\documentclass{aa}  

\usepackage{txfonts}
\usepackage[pdfpagelabels=false]{hyperref}      
\hypersetup{colorlinks=true,linkcolor=blue,citecolor=blue,filecolor=blue,urlcolor=blue,}
\usepackage{placeins}    
\usepackage{graphicx}	
\usepackage{amsmath}	
\usepackage{amssymb}	
\usepackage{booktabs}	
\usepackage{xspace}     
\usepackage[version=4]{mhchem}
\usepackage{ulem}
\usepackage{caption}
\usepackage{subcaption}
\usepackage{lastpage}
\usepackage[rightcaption,ragged]{sidecap} 
\sidecaptionvpos{figure}{c}
\makeatletter
\renewcommand*\aa@pageof{, page \thepage{} of \pageref*{LastPage}}
\makeatother
\renewcommand\makeLineNumber{}

\begin{document}

   \title{Chemical transformation of CO in evolving protoplanetary discs across stellar masses: a route to C-rich inner regions}

   \titlerunning{CO chemical transformation and inner disc C/O across stellar mass}

   \author{Andrew D. Sellek
          \inst{1}
          \and
          Ewine F. van Dishoeck
          \inst{1,2}
          }

   \institute{Leiden Observatory, Leiden University, 2300 RA Leiden, The Netherlands\\
   \email{sellek@strw.leidenuniv.nl}
   \and
   Max-Planck Institut f\"{u}r Extraterrestrische Physik (MPE), Giessenbachstr. 1, 85748, Garching, Germany}

   \date{Received 17 April 2025; accepted 14 July 2025}

  \abstract
   {Protoplanetary discs around Very Low Mass Stars (VLMS) show hydrocarbon-rich MIR spectra indicative of C/O>1 in their inner discs, in contrast to those around higher-mass hosts which are typically richer in O-bearing species.}
   {The two scenarios proposed to elevate C/O around the inner discs of VLMSs are the release of C by eroding carbonaceous grains, or the advection of O-depleted gas from the outer disc. However, if CO gas remains abundant, sufficiently O-depleted material cannot be produced. We test whether the chemical transformation of CO into other species allows the transport scenario to produce C/O significantly in excess of 1.}
   {We track the inner disc C/H and O/H over time using a 1D disc evolution code that models the transport of gas and ice phase molecules and includes the conversion of some species into others to represent key reaction pathways operating in the midplane. We explore the influence of disc mass, size, ionization rate, and the presence of a dust trap.}
   {The inner disc C/O increases over time due to sequential delivery where O-rich species (e.g. \ce{H2O}) give way to C-rich species (e.g. \ce{CH4}). To reach C/O>1, separating C and O is key, hence the gas phase destruction of CO by \ce{He+}, which liberates C, is critical. 
   Ionization drives the midplane chemistry and must have rates $\gtrsim\!10^{-17}\,\mathrm{s^{-1}}$ (at least for VLMSs) for significant chemical evolution within the disc lifetime. However, the rates must be $\lesssim\!10^{-17}\,\mathrm{s^{-1}}$ for T Tauri stars to ensure their C/O remains <1 for the first few Myr.
   Initially more compact discs lose O-rich ices faster and reach a higher C/O.
   A warm dust trap between the \ce{CH3OH} and \ce{CH4} snowlines traps \ce{CH3OH} ice (formed via hydrogenation of CO ice) for long enough to be photodissociated, providing an alternative way to liberate the C that started in CO in the form of \ce{CH4} gas that keeps the inner disc significantly C-rich.}
   {Destruction of gaseous CO, combined with gas advection and radial drift, can deplete O enough and produce sufficient hydrocarbons to explain the typical C/O>1 of VLMSs. While their C/O is typically higher than for T Tauri stars due to the faster sequential delivery, achieving values significantly in excess of 1 likely also requires higher ionization rates and more compact discs than for T Tauri stars. Observations of older discs may distinguish whether a higher ionization rate is indeed required or if the faster physical evolution timescales alone are sufficient.}


   \maketitle
%

\section{Introduction}
\vspace{-10pt}
The composition of dust and gas in protoplanetary discs determines that of any planets forming therein.
Consequently, elemental abundances such as C/H and O/H, and particularly ratios such as C/O, have been suggested to trace the formation location and mechanisms: at overall solar composition, the C/O ratio of both the solids and gas in a static disc would rise with distance from the central star as successive snowlines of \ce{H2O}, \ce{CO2}, and \ce{CO} are crossed \citep{Oberg_2011}.
However, there is a developing understanding that the picture is more complex than implied by such static models. Disc evolution models show significant redistribution of molecules resulting from the rapid radial drift of dust grains, sublimation of molecular ices as these grains cross snowlines, and the subsequent slow accretion of this enriched gas inwards \citep{Oberg_2016,Booth_2017,Krijt_2018}. Further complications include evolution of the disc temperature causing the snowlines to migrate \citep{Miley_2021}, migration of planets themselves \citep{Turrini_2021,Pacetti_2022}, trapping of volatile species in less-volatile ices \citep{Collings_2004,Ligterink_2024}, vertical temperature gradients that lead to `snow surfaces', and chemical evolution of the gas and ice \citep{Eistrup_2023}.

Confirming how abundances vary across a disc thus requires comparing to measurements from observations.
While the Atacama Large Millimeter/submillimeter Array (ALMA) is sensitive to material in the outer disc red($\sim$10s au) \citep[e.g.][]{Kama_2016,Bergner_2019,Miotello_2019,Oberg_2021,Sturm_2022,Bergin_2024}, observations in the infrared, such as with JWST's Mid-Infrared Instrument (MIRI) Medium Resolution Spectroscopy (MRS) mode, are better suited to characterise warm gas in the inner disc ($<$ few au).

So far, the inner disc chemistry depends strongly on stellar mass. 
In discs around stars $\gtrsim\!0.3\,M_\odot$ (T\,Tauri stars, hereafter TTS), the most frequently detected molecules are \ce{H2O}, \ce{CO2}, \ce{C2H2} and \ce{HCN} \citep{Grant_2023,Grant_2024,Banzatti_2023,Banzatti_2025,Gasman_2023,Gasman_2025,Henning_2024,Pontoppidan_2024,RomeroMirza_2024,Schwarz_2024,Temmink_2024a,Salyk_2025,Vlasblom_2025,Arulanantham_2025}, with retrieved emitting areas that imply emission mostly inside $\lesssim 1\,\mathrm{au}$. This pattern, with strong O-bearing species, is thought to occur when O is in excess with respect to C, i.e. C/O<1.
Conversely, discs around very low mass stars (VLMS) $\lesssim\!0.3\,M_\odot$ have stronger \ce{C2H2} relative to the other molecules \citep{Grant_2025}, strengthening earlier Spitzer analyses \citep{Pascucci_2009b,Pascucci_2013}, with MIRI-MRS revealing and resolving a rich hydrocarbon chemistry with orders of magnitude larger column densities than thought before \citep{Tabone_2023,Arabhavi_2024,Kanwar_2024b,Arabhavi_2025a}.
Thermochemical models imply that C/O>1 is required to achieve such conditions \citep{Najita_2011,Kanwar_2024b}, which are found at a wide range of ages \citep{Long_2025}.
Notable exceptions to this pattern are DoAr\,33 \citep{Colmenares_2024}, a $1.1\,M_\odot$ star with a hydrocarbon-rich chemistry consistent with C/O=2-4, and the \ce{H2O}-rich inner disc of the VLMS Sz\,114 \citep{Xie_2023}.

Two broad explanations have been proposed for the high C/O cases \citep{Tabone_2023,Arabhavi_2024}. On the one hand, C/H could be increased by the destruction or erosion of a refractory carbon reservoir. Conversely, the O/H could be depleted in the inner disc gas as a result of the chemical evolution expected due to transport \citep{Mah_2023} or retention of O-rich ices in dust traps or planetesimals \citep{Najita_2013,McClure_2020}. Distinguishing between these is important as they predict distinct C/H and O/H for a given C/O.

In the erosion scenario, refractory C could be liberated at high temperatures by pyrolysis of organic solids (their thermal decomposition at $\gtrsim\!500\,\mathrm{K}$) at the `soot line' \citep{Kress_2010,LiBergin_2021,Houge_2025b}, or chemical erosion such as oxidation reactions \citep{Lee_2010} or hydrogen chemisputtering \citep{Lenzuni_1995,Borderies_2025}. The sublimation of amorphous C is less likely as it occurs at considerably higher temperatures \citep[$\gtrsim\!1100\,\mathrm{K}$,][]{Gail_2017}. Alternatively, the radiation field can lead to destruction of grains or PAHs by ultraviolet photons \citep[photolysis,][]{Anderson_2017,Bosman_2021b,Vaikundaraman_2025} or X-rays \citep{Siebenmorgen_2010}.

In the O-depletion scenario, ices desorbing from inwardly drifting dust deliver less-volatile, more oxygen-rich molecules closer to the star.
Consequently, the inner disc is expected to first go through a high O/H, low C/O, \ce{H2O}-enriched phase. This is followed by an increase in C/O as the \ce{H2O} accretes onto the star and successively carbon-rich molecules - \ce{CO2}, and later \ce{CO} and \ce{CH4} - arrive \citep{Mah_2023,Mah_2024,Sellek_2025}; the strength and duration of these phases can depend on the formation, location, and leakiness of any dust traps that block drift.
Recent observations of VLMSs suggest that the inner disc may indeed become increasingly carbon rich when comparing older discs to younger ones \citep{Long_2025,Arabhavi_2025a}.

Crucially, \citet{Mah_2023} argue that the colder temperatures of discs around VLMSs result in the snowlines lying closer to the star and so a faster progression through this evolutionary pattern. Thus at a given age, discs around VLMSs are likely to show a higher C/O.
The disc transport models of \citet{Mah_2023} do cross C/O=1 (required for rapid hydrocarbon formation) but only due to their inclusion of abundant \ce{CH4} (as a species containing C but not O). However, those transport models never suggest a C/O>1.5, despite such values being inferred for some discs \citep{Colmenares_2024,Kanwar_2025,Long_2025}. This is despite making generous assumptions for the initial abundance of \ce{CH4} in the disc \citep[see also discussion in][]{Houge_2025b} that vastly exceed the amount inherited from ices in protostellar envelopes \citep{Oberg_2011b,Boogert_2015} or found in cometary ices \citep{Mumma_2011,Altwegg_2019,Schuhmann_2019}.
It thus remains unclear whether transport alone can indeed explain the difference in C/H, O/H, and C/O between VLMSs and TTSs, or if in situ C/H enhancement is required.

The fundamental barrier to achieving a high C/O ratio is that it is buffered by CO being the most abundant C-bearing molecule in these disc models. With C atoms thus always being associated with around one O atom, transport alone is unable to avoid C/O$\sim\!1$.
Raising C/O further thus requires destroying CO in a way such that the C atoms are liberated in the process.

It is indeed often found that CO and its isotopologues - as observed with ALMA - are less abundant in the outer disc around TTSs than the ISM, typically by factors 10-100 \citep{Favre_2013,Kama_2016b,McClure_2016,Anderson_2022,Trapman_2022,PanequeCarreno_2025}. This so-called CO `depletion' sets in on timescales of $\sim$\!1\,Myr \citep{Harsono_2014,Zhang_2020b,Bergner_2020}. While some effect is expected due to freeze-out and photodissociation, the term `volatile CO depletion' refers to a lack of CO on top of these processes \citep{Miotello_2017}.
Models have explored both physical routes - involving the sequestering of CO ice onto pebbles that settle to the midplane and drift \citep{Kama_2016b,Krijt_2020,Powell_2022} - and chemical routes - involving the conversion of CO molecules to less volatile species \citep{Yu_2016,Molyarova_2017,Bosman_2018b,Schwarz_2019,Krijt_2020,vanClepper_2022,Zwicky_2025} - to achieving this depletion\footnote{Some other works use `depletion' only if CO is less abundant than predicted given chemical transformation processes as well as freeze-out and photodissociation, which has lead to competing claims \citep[e.g.][]{Ruaud_2022,Pascucci_2023b} of whether CO is depleted \citep[see discussion in Section 4.1 of][]{PanequeCarreno_2025}. In the works which find CO `depletion', grain surface processes are not modelled in detail so CO remains abundant at the $10^{-4}$ level; the volatile C abundance is instead artificially reduced in order to approximate the transformation of CO to less volatile species and thus reduce the CO abundance. Ultimately, all works agree on the need to transform CO to less volatile species in order to reproduce the observed line fluxes.}.
For our purposes, however, physical sequestration does not liberate C, so we must also consider chemical routes.

Conversely, liberating C from CO cannot by itself alter C/O (or lead to C/O>1) if the liberated O is not removed.
While other works also argue for the importance of liberating the C from CO in order to produce hydrocarbons, for example as a result of ionizing X-rays \citep{Walsh_2015}, they employ physically static models.
However, any O-bearing species produced are generally less-volatile than the hydrocarbons. Thus, the differential transport of vapour and ice due to dust radial drift, which is not considered in the thermochemical models, could separate the liberated C and O reservoirs and produce conditions where C/O>1.

In this work, we implement simple prescriptions for chemical transformation of CO in a disc evolution code \citep{Booth_2017}, as described in Section \ref{sec:methods}, in order to study whether this can prevent CO transport to the inner disc in a way that results in C/O>1. We thus explore when depletion of O as a result of global-scale disc evolution could explain high C/O measurements in the inner disc via grids of such models.
Section \ref{sec:results} presents the evolution of C/H, O/H, and C/O for these grids, while Section \ref{sec:discussion} considers which conditions would need to be met for discs around VLMSs to have higher C/O than those around TTSs and how to go about determining if CO-depleted outer disc gas is indeed responsible for the elevated C/O.
We summarise our conclusions in Section \ref{sec:conclusions}.

\section{Methods}
\label{sec:methods}
\subsection{Disc evolution model}
We use the 1D Disc Evolution code by \citet{Booth_2017}, including the updates to the dust transport in \citet{Sellek_2020b} and the chemistry in \citet{Sellek_2025}, to model the evolution of discs around $1\,M_{\odot}$ and $0.1\,M_{\odot}$ stars (Table \ref{tab:parameters}).
We model the discs for 10 Myr, which spans the ages of the nearby star-forming regions studied in Cycles 1 \& 2 with JWST, including the older Upper Sco region; the endpoint must be defined manually since we do not include any process to disperse the discs \citep[see][for models including photoevaporation]{Lienert_2024,Lienert_2025}.
In brief, the code computes the viscous evolution of the disc's gas according to the \citet{Shakura-Sunyaev_1973} $\alpha$ prescription where $\nu=\alpha c_{\rm S} H$ for sound speed $c_{\rm S}$ and scale height $H$ and we assume $\alpha=10^{-3}$.
On top of this, the dust evolution is computed using the two-population model \citep{Birnstiel_2012} implemented following \citet{Laibe_2014,Dipierro_2018}. In this approach, the majority of mass is contained in large grains, which have sizes limited by either drift or fragmentation and which undergo significant radial drift, while there is always a remnant population of small grains which remain coupled to the gas. The grains are assumed to fragment at a velocity of $1\,\mathrm{m\,s^{-1}}$ for bare dust grains and $10\,\mathrm{m\,s^{-1}}$ for icy dust grains.

Several gas and ice phase molecular tracers are advected following the motion of the gas or dust components respectively. At each time step, the equilibrium between freeze-out and thermal desorption is calculated to determine the fraction of each molecule in the gas and ice phases.
The species included in this work are those used by \citet{Sellek_2025} (\ce{H2O}, \ce{CO2}, \ce{CO}, \ce{CH4}, \ce{CH3OH}, \ce{O2}, C-grains, and Si-grains), plus we additionally investigate \ce{C2H2}, \ce{C2H4}, and \ce{C2H6}.
All prefactors and binding energies used \citep[which for the hydrocarbons are for pure multilayer ices from][]{Behmard_2019}, as well as resulting snowline locations for each $M_*$, are given in Table \ref{tab:binding}.

For the initial conditions, we assume a surface density profile given by \citep{Lynden-Bell_Pringle_1974}:
\begin{equation}
    \Sigma_{t=0}(R) = \frac{M_{\rm D,0}}{2\pi R_{\rm C,0}} \frac{1}{R} \exp\left(-\frac{R}{R_{\rm C,0}}\right)
    \label{eq:LBP}
    ,
\end{equation}
where we explore several values for the initial mass $M_{\rm D,0}$ and radius $R_{\rm C,0}$ (Table \ref{tab:parameters}).
All our $M_{\rm D,0}$ are $\lesssim0.1M_*$ so represent discs in the gravitationally stable regime. This regime may follow a period of gravitational instability maintained by infall, where non-axisymmetric substructures and temperature profiles that cannot be captured in 1D form; although these effects can already leave an imprint in the distributions of different molecules \citep{Molyarova_2025}, we use uniform initial abundances (Appendix \ref{appendix:chemistry}) in this work.
The $R_{\rm C,0}$ follow the range of values inferred by \citet{Trapman_2023} from CO gas disc radii and are used for both $M_*$ investigated.
The $\Sigma(R)$ profiles for $t=0\,1\,\&\,5\,\mathrm{Myr}$, are shown for fiducial parameters in Fig. \ref{fig:zetaCR}.

The temperature is prescribed as a power-law proportional to $R^{-1/2}$ and scaled based on the aspect ratio at 1\,au ($h_0$), where we assume $h_0\propto M_*^{-0.425}$ \citep{Sinclair_2020}:
\begin{equation}
    T_0 = \frac{GM_*\mu}{R\mathcal{R}}h_0^2
    .
\end{equation}
This profile is assumed constant: we do not model the effects of stellar evolution \citep[unlike e.g.][]{Miley_2021} or heating from hydrodynamic processes in the disc \citep[unlike e.g.][]{Molyarova_2025} - and applies to both dust and gas, since they couple thermally at midplane densities.

While Equation \ref{eq:LBP} defines a ``smooth'' density profile that is free of substructures or dust traps, we also consider perturbations to the gas surface density. These were introduced by varying the viscosity locally, centred around a radius $R_{\rm gap}$ and with an amplitude $A_{\rm gap}$,
where the perturbation is fully developed by time $t_f$ \citep[see][for further details]{Sellek_2025}.
We choose include either $R_{\rm gap}=5$\,au, which creates a `warm' trap, or $R_{\rm gap}=50$\,au, leading to a `cold' trap. These are so called because they lie at warmer and colder temperatures than the freeze-out of CO and \ce{CH4}; discs have been observed to show differences in the strength of their \ce{C2H} emission, related to their C/O ratio, between these cases \citep{vanderMarel_2021d}.
While the most easily resolved substructures lie at 10s au \citep{Bae_2023}, high-resolution surveys have shown they can form at distances as close as 5\,au in a small fraction ($\sim\!10\%$) of compact discs \citep{GuerraAlvarado_2025}.
Moreover, we model cases where traps a) are present from the start ($t_f=0$), which maximises their chances of impacting pebble drift and volatile delivery \citep{Mah_2024,Sellek_2025}, or b) only form by $t_f=1$\,Myr, as they are more rarely observed in the earlier embedded stages \citep{Ohashi_2023,Hsieh_2024}. This lack of substructures at earlier times may be due to observational limitations or the fact that planets (if they are indeed the origin of traps) may take Myr timescales to form and grow to the pebble isolation mass, especially as it is harder for them to open a gap in younger, warmer, discs \citep{Nazari_2025}.

\begin{table}[t]
    \centering
    \caption{Parameters explored in our model grids. Fiducial values are highlighted in bold.}
    \begin{tabular}{c|c|c}
    \hline\hline
        Parameter & Value in $1\,M_{\odot}$ models & Value in $0.1\,M_{\odot}$ models \\
    \hline
        $R_{*}$             & 2.5 $R_{\odot}$ & 1.0 $R_{\odot}$ \\
        $h_{0}$             & 0.021 & 0.056 \\
        $M_{\rm D,0}$       & 0.025, $\mathbf{0.05}$, 0.1\,$M_\odot$ & 0.0025, $\mathbf{0.005}$, 0.01\,$M_\odot$ \\
        $R_{\rm C,0}$       & 12, 16, 25\,au              & 12, 16, 25\,au \\
    \hline
        $\alpha$            & \multicolumn{2}{c}{$10^{-3}$} \\
        $\zeta_{\rm \ce{H2}}$   & \multicolumn{2}{c}{$1.3 \times 10^{-18}$, $\mathbf{1.3 \times 10^{-17}}$, $1.3 \times 10^{-16}$\,$\mathrm{s^{-1}}$} \\
    \hline
        $A_{\rm gap}$       & \multicolumn{2}{c}{$\mathbf{0}$, 10} \\
        $R_{\rm gap}$       & \multicolumn{2}{c}{$\mathbf{5}$ au, 50\,au (for $R_{\rm C,0}=50\,\mathrm{au}$ only)} \\
        $t_{\rm f}$         & \multicolumn{2}{c}{0, 1\,Myr} \\
    \hline
    \end{tabular}
    \label{tab:parameters}
\end{table}

\subsection{Reactions}
\begin{figure}
    \centering
    \includegraphics[width=\linewidth]{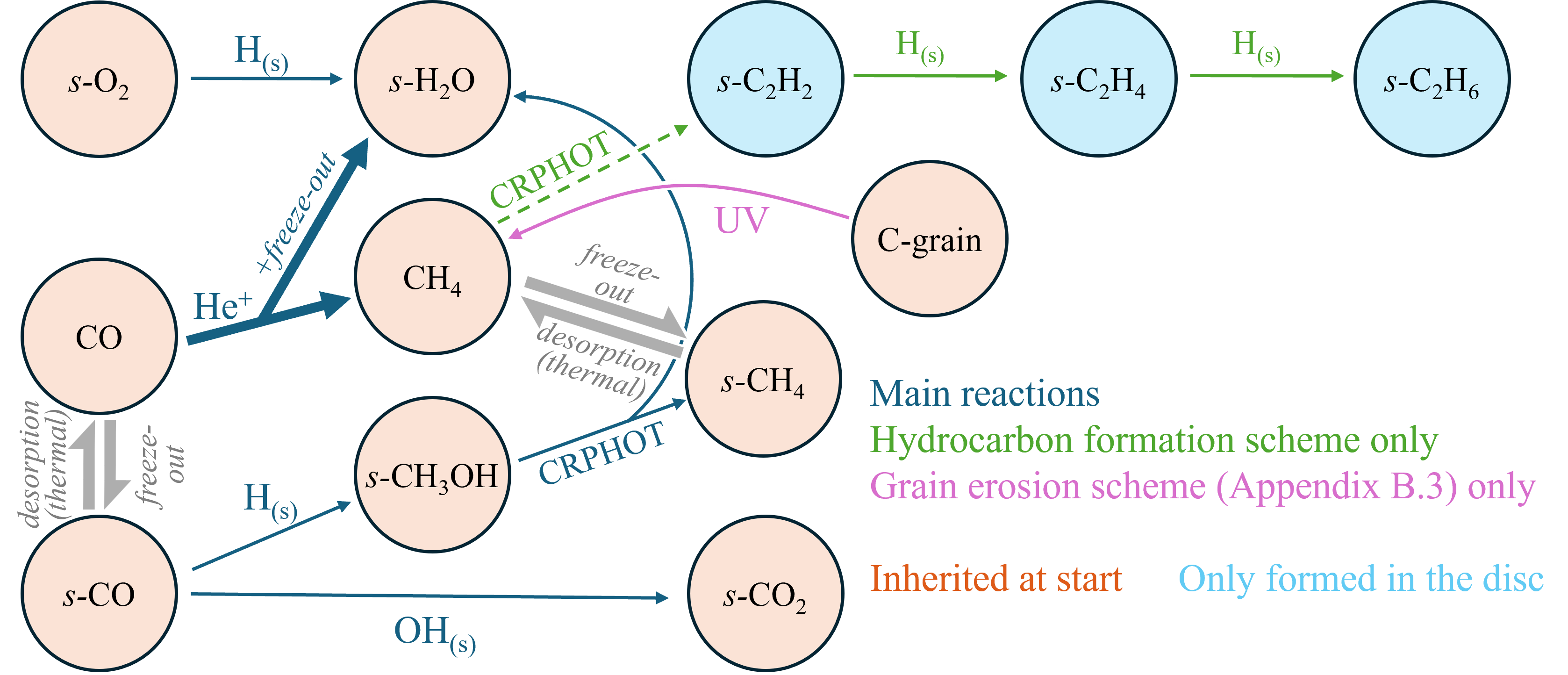}
    \caption{Summary of the reaction scheme(s) used in this work. '\textit{s-}' indicates that the species is in the ice phase. The process in bold is found to be most critical to raising the inner disc C/O.}
    \label{fig:reactionScheme}
\end{figure}

The chemistry in disc midplanes is driven by ionisation (Sections \ref{sec:methods_ionisation}, \ref{sec:discussion_ionisation}), often assumed to result from cosmic rays (CRs), which also produce a secondary UV field \citep{Prasad_1983}.
Molecules can be photodissociated by this UV  - or react with the ions or radicals created by the CRs or secondary photons - with rates that imply changes on Myr timescales for typical ionisation rates.
We do not include a detailed chemical network, but instead a toy model based on results from thermochemical disc models \citep[e.g.][]{Walsh_2015,Yu_2016,Eistrup_2016,Eistrup_2018,Bosman_2018b}.
Our approach is summarised in Figure \ref{fig:reactionScheme} and involves simple prescriptions for the net effect of major chemical pathways, based on the initial rate-limiting steps, similar to that used by \citet{Krijt_2020}.

For each time step of the disc evolution code (typically a few years), the rates are calculated and the abundances of molecules updated in an Eulerian manner; the time step for the disc evolution is taken as the minimum of the shortest time step required to model the transport processes and the shortest timescale required to deplete any molecular tracer by a factor $e$ in order to ensure that no molecular abundance ever becomes $\leq0$.

\subsubsection{CO transformation}
\citet{Bosman_2018b} characterised CO destruction in disc midplanes by three main routes; in terms of the net consumption of our molecular reservoirs, these are 
\begin{enumerate}
    \item Dissociative ionization of gas phase CO by He\textsuperscript{+}, where we assume that the liberated C\textsuperscript{+} reacts to form \ce{CH4}
    \vspace{-6pt}
    \begin{equation}{\ce{CO + 3 H2 ->[He^+] CH4 + H2O}}\end{equation}
    \item Hydrogenation of CO ice on grain surfaces, which forms \ce{CH3OH} via HCO
    \vspace{-6pt}
    \begin{equation}{\ce{CO(s) + 2 H2 ->[H] CH3OH(s)}}\end{equation}
    \item Reaction of CO with OH on grain surfaces to form \ce{CO2}
    \vspace{-6pt}
    \begin{equation}{\ce{CO(s) + H2O(s) ->[OH] CO2 + H2}}\end{equation}
\end{enumerate}
Above each arrow are intermediaries, created by ionization or photodissociation, required for the reactions (Appendix \ref{appendix:chemistry}).

We treat grain-surface reactions as tunnelling processes following \citet{Hasegawa_1992}, where the top two surface layers are active \citep{Bosman_2018b}. H is assumed to dominate the diffusion, at the maximum of the thermal hopping and tunnelling rates, for reactions it participates in.
The barrier to diffusion is set as $0.3$ of the binding energy \citep[for H we take the amorphous solid water value in][]{Minissale_2022}.
The barrier width for the reactions is taken to be the standard 1\,\r{A} and the barrier heights are equal to those in gas-grain codes used for the thermochemical modelling of protoplanetary discs \citep[][see Table \ref{tab:barriers}]{Walsh_2015}. 
The 300\,K  barrier for \ce{O2 + H -> HO2} used in that network \citep{Lamberts_2013,Taquet_2016} results in a very short reaction timescale. For computational efficiency we thus assume that any frozen-out \ce{O2} is entirely consumed (otherwise this process seriously throttles the code time step). We checked that this assumption had negligible impact on our results: the reaction rate of \ce{O2} is fast enough that in either case, very little \ce{O2} survives for the Myr timescales that we are interested in.
Gas-phase reaction rates are taken from the \textsc{umist rate22} database \citep{Millar_2024}.

\subsubsection{Further reactions}
\ce{CH3OH} ice, as formed by hydrogenation of CO, may undergo UV photodissociation.
For simplicity, we follow \citet{Krijt_2020} and assume that this always splits the C-O bond, with the \ce{CH3} and O being rapidly hydrogenated to \ce{CH4} and \ce{H2O}:
\begin{equation}
{\ce{CH3OH(s) + H2 ->[UV] CH4(s) + H2O(s)}}.
\end{equation}
We note however, that the branching ratios for photodissociation of \ce{CH3OH} are uncertain: other works \citep{Laas_2011,Drozdovskaya_2014} adopt a `standard' assumption of 60\% \ce{CH3} production, while laboratory experiments suggest it could be even lower \citep[$\lesssim\!14\%$][]{Oberg_2009d}; in this sense, our rate is an upper limit for the conversion of \ce{CH3OH} ice to \ce{CH4}.

Moreover, the most commonly observed hydrocarbon in the inner regions of protoplanetary discs is \ce{C2H2}. We assume this forms after photodissociation of gas-phase \ce{CH4}
\begin{equation}
{\ce{CH4 ->[UV] 1/2 C2H2 + 3/2 H2}}.
\end{equation}
\ce{C2H2} may be destroyed by hydrogenation on grain surfaces \citep{Aikawa_1999,Yu_2016,Bosman_2018b}. The radicals \ce{C2H3} and \ce{C2H5} are highly reactive with no barrier for hydrogenation \citep{Hasegawa_1992} and hence very short lived. Thus we consider only the more stable \ce{C2H4} and \ce{C2H6}.
\begin{align}
    \ce{C2H2 + H2 &->[H] C2H4} \\
    \ce{C2H4 + H2 &->[H] C2H6}.
\end{align}
The initial abundances of these hydrocarbons were assumed to be 0 everywhere so we could focus on their production routes.

Another way to liberate additional C is by destroying or eroding carbonaceous grains. In Appendix \ref{appendix:photolysis} we estimate UV photolysis of grains in the outer disc to have little effect on our models, and so henceforth neglect this process in this work. We refer the reader to \citet{Houge_2025b} for an exploration of the role of thermal decomposition of grains in the inner disc.

\subsubsection{Ionization Rates}
\label{sec:methods_ionisation}
Every process discussed above depends indirectly on the ionization rate. At the midplane, CRs are often assumed to be the principal source of ionization - with a fiducial rate assumed in databases such as \textsc{umist} \citep{Millar_1991,Millar_2024} of $\zeta_{\rm \ce{H2}}=1.3\times10^{-17}\,\mathrm{s^{-1}}$ \citep[a dense cloud value;][]{Prasad_1980} - and thus drive the chemistry \citep{Eistrup_2016,Eistrup_2018}.
CRs are subject to attenuation, with an average penetration depth of $\Sigma_{\rm CR} = 96\,\mathrm{g\,cm^{-2}}$ \citep{Umebayashi_1981}; for simplicity, we assume they are attenuated by the vertical column of gas\footnote{If CRs detour along sheared magnetic field lines, the surface density they traverse to reach the midplane at a distance $R$ can be greater than $\Sigma(R)$ by around 2 orders of magnitude \citep{Fujii_2022}}.

When CRs are highly attenuated, ionization does not cease entirely. Short-lived radionuclides (SLRs) can provide ionization rates of up to $7.6\times10^{-19}$ \citep{Umebayashi_2009}, though this value will decay over time as the SLRs have half-lives of around 1\,Myr \citep[e.g.][]{Cleeves_2014b,Eistrup_2018}. This rate is too low to produce much chemical evolution within typical protoplanetary discs lifetime \citep{Eistrup_2018} but we include it as a floor value nonetheless.

The ionizing particle flux incident on a disc can also vary from the fiducial value. For example, the flux of galactic CRs reaching Earth is modulated by solar magnetic activity; the increased activity of young stars may reduce the ionization rate to as little as $\sim\!10^{-20}\,\mathrm{s^{-1}}$ even before attenuation in the disc \citep{Cleeves_2013}.
Conversely, energetic particles may be accelerated locally by flares from the central star \citep{Feigelson_2002,Rab_2017,Rodgerslee_2017,Rodgerslee_2020}, reconnection events in the interaction region between the stellar and disc magnetic fields \citep{Orlando_2011,Colombo_2019,Brunn_2023}, or shocks in protostellar jets \citep{Padovani_2015,Padovani_2016}.
Finally, hard X-rays also produce ionization, with midplane ionization rates typically of the order of $10^{-20}-10^{-19}\,\mathrm{s^{-1}}$ \citep{Cleeves_2013,Cleeves_2014b,Rab_2017}.
To encompass these possibilities in a general way, agnostic to the origin of the variations, we define a single ionization rate incident on the disc $\zeta_{\rm \ce{H2}}$ and explore values raised or lowered uniformly by a factor 10 with respect to the fiducial CR value (Table \ref{tab:parameters}).

\section{Results}
\label{sec:results}
\subsection{No C-chemistry}
\begin{figure*}[t]
\centering
\begin{subfigure}[t]{0.48\linewidth}
    \centering
    \includegraphics[width=1.05\linewidth]{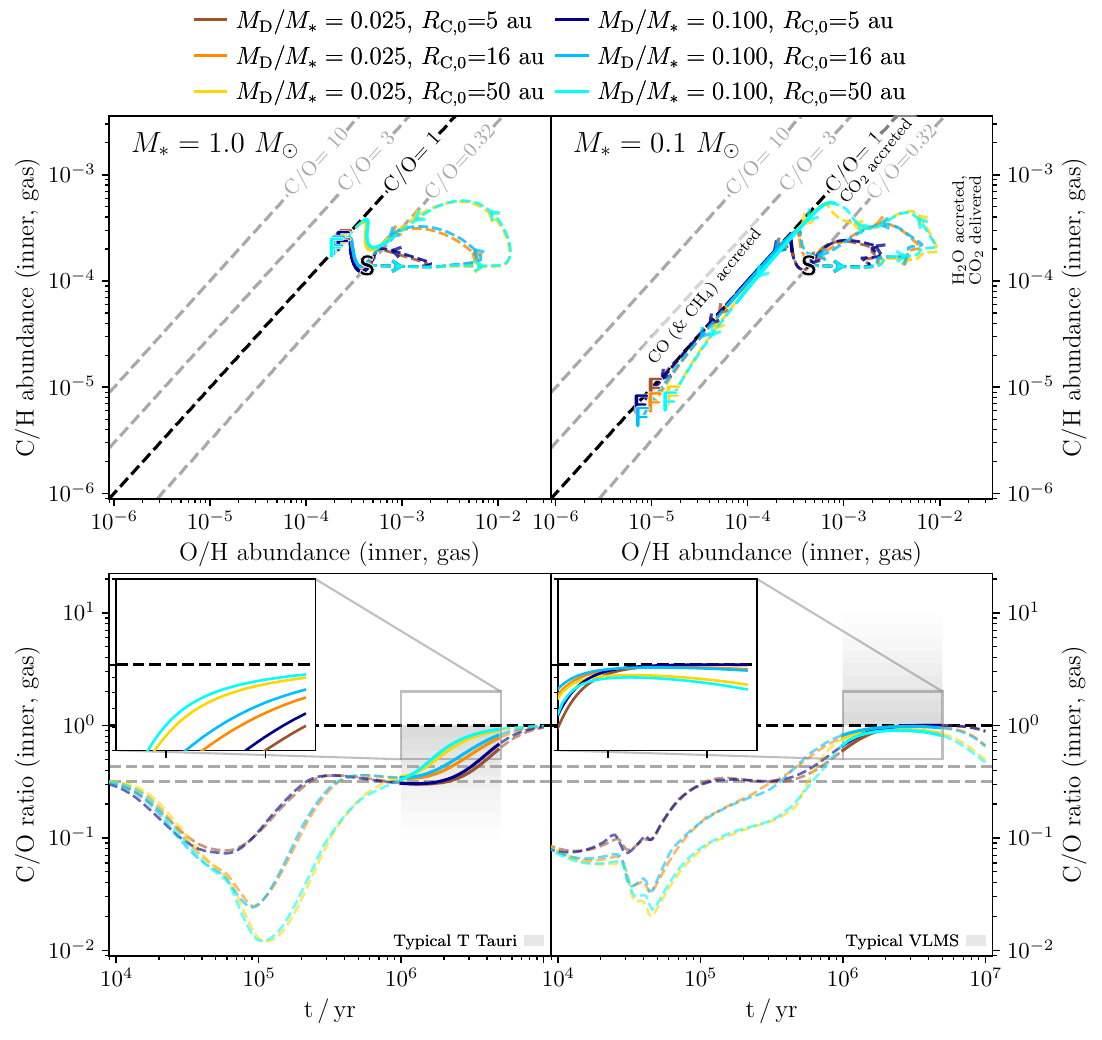}
    \vspace{-18pt}
    \caption{}
    \label{fig:noConv}
\end{subfigure}
\begin{subfigure}[t]{0.48\linewidth}
    \centering
    \includegraphics[width=1.05\linewidth]{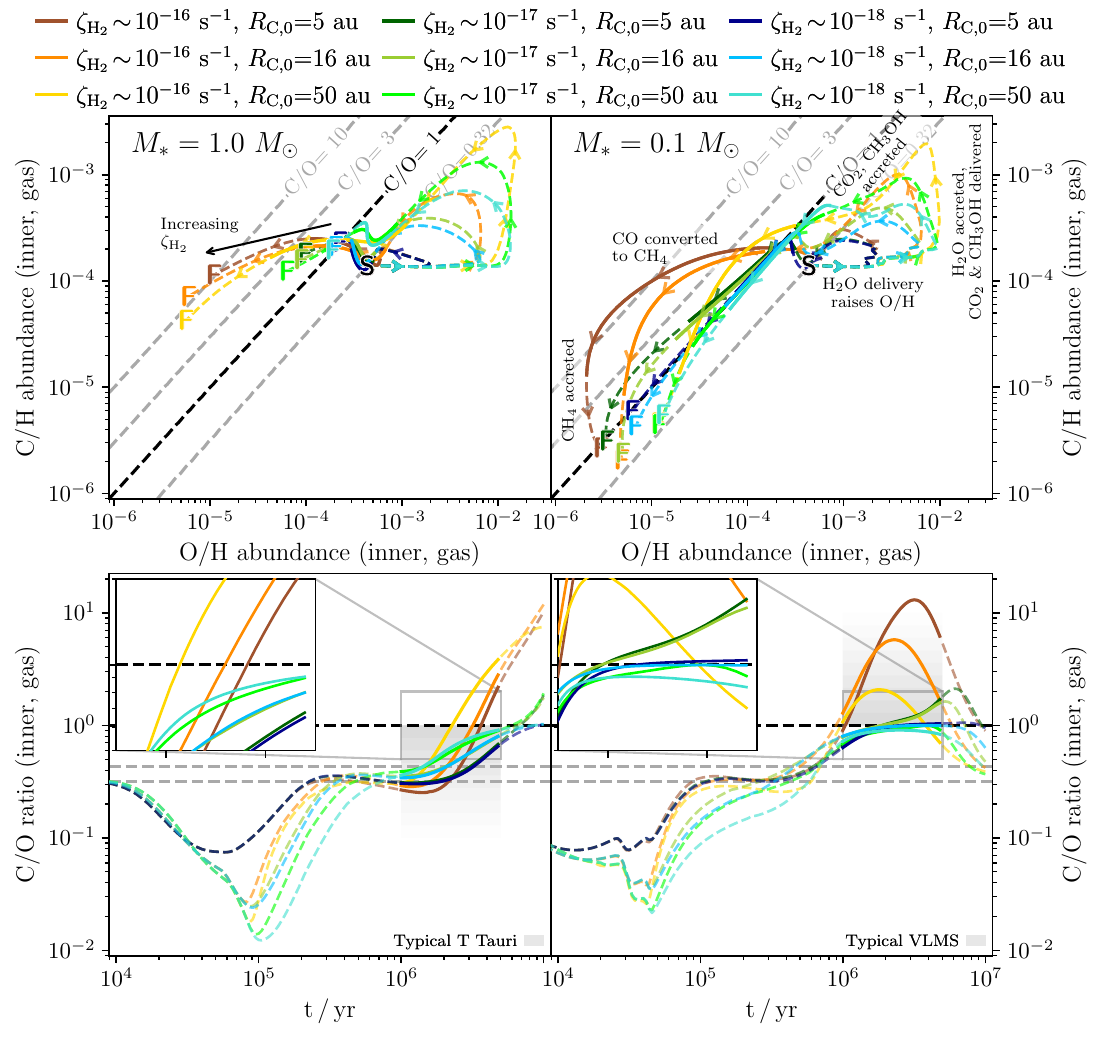}
    \vspace{-18pt}
    \caption{}
    \label{fig:COdepletion}
\end{subfigure}
    \vspace{-3pt}
\caption{
    10\,Myr of evolution of different disc models: a) models that do not include reactions of C-bearing species for various initial disc-to-star mass ratios and initial disc sizes; b) models that include reactions of CO, all with a fixed initial disc-to-star mass ratio of 0.05 but different combinations of ionization rate and disc size.
    In each panel, the tracks start at 'S' and evolve in the direction of the arrows, ending at 'F'; the solid sections represent ages of 1-5\,Myr appropriate to most discs studied so far with MIRI-MRS.
    In each block, the left-hand columns show models around a 1.0 $M_{\sun}$ star, representative of TTSs, while the right-hand columns show models around a 0.1 $M_{\sun}$ star, representative of VLMS.
    The top rows plot the discs in the plane of the inner-disc gas-phase C/H vs gas-phase O/H. Dashed lines indicate where C/O=0.32,1,3,10.
    The bottom rows plot the gas-phase C/O in the inner disc over time. The horizontal dashed lines show ratios of 0.32 (initial ratio of volatile abundances), 0.43 (ratio of total abundances) and 1 and the lightly shaded regions indicate the typically observed range of C/O at 1-5\,Myr for each stellar mass regime.
}
\end{figure*}

As a baseline, we consider a model set without any reactions of C-bearing species; the models only include the oxidation of \ce{O2} ice to \ce{H2O} (otherwise \ce{O2} could provide a large, unconstrainable
volatile-O reservoir). Figure \ref{fig:noConv} shows the evolution of these models in the plane of the gas-phase C/H and O/H in the inner disc and the evolution of the resulting C/O over time. We measure this at the inner boundary of our disc model (0.1 au), but it should be representative of the conditions inside the \ce{H2O} snowline. We shade regions where C/O<1 for TTSs and C/O>1 for VLMSs to illustrate the typical constraints inferred without placing any weight on exact individual ages or C/O values, which vary depending on the method used to derive them.

Figure \ref{fig:noConv} shows that initially O/H is enhanced (with a corresponding drop in C/O) as radial drift delivers O-rich \ce{H2O} ice to the inner disc. As the \ce{H2O} vapour is accreted onto the star, lowering O/H again, \ce{CO2} becomes the dominant molecule in the inner disc \citep{Sellek_2025}, leading to an increase in C/H and an overall rise in C/O. The initial disc size $R_{\rm C,0}$ has a small effect on the evolution. For more extended discs, a greater fraction of the molecular budget is initially frozen out, resulting in
slightly
larger enhancements in these phases. The evolution happens faster for lower-mass stars, reflecting the snowlines lying closer to the central star \citep{Mah_2023}. The initial disc-to-star mass ratio has negligible effect.

Over time, CO becomes the dominant molecular species, and
the C/O reaches $\sim\!1$. This is the end state after 10\,Myr for $1.0\,M_{\odot}$
stars, while for VLMSs, the closer-in snowlines result in enough
time to deplete the CO vapour significantly through accretion
onto the star, resulting in lower final C/H and O/H. At late times,
a small number of dust grains, which originate at large radii and
never grew large enough to undergo significant radial drift,
remain in the disc. These, by definition, experience the least decoupling from the gas and so produce little segregation of C- and O-bearing molecules. Consequently, after the remnants of the earlier phases of strong radial drift completely accrete onto the star (only for VLMS discs does this happen within 10\,Myr), these small grains reset the elemental composition to the initial volatile C/O=0.32 assumed in the models. A gradual fall in C/O towards this value is thus seen at the latest times $\gtrsim\!7\,\mathrm{Myr}$.
Other than in this very last phase, at a given time the VLMS discs show higher C/O as a result of their faster evolution \citep{Mah_2023}.

The C/O of the disc can be calculated as
\begin{equation}
    \frac{\rm C}{\rm O} = \frac{\rm CO+CH_4+CO_2+CH_3OH}{\rm CO+2\times O_2+2\times CO_2+CH_3OH+H_2O}
    .
\end{equation}
In the absence of all O-bearing species other than CO, which all have shorter lifetimes in the disc than CO (including \ce{O2} due to the efficient oxidation to \ce{H2O}), this reduces to
\begin{equation}
    \frac{\rm C}{\rm O} = \frac{\rm CO+CH_4}{\rm CO}
    .
\end{equation}
As \ce{CH4} is only marginally less volatile than CO, their ratio is only weakly altered by the relative dust and gas evolution. Thus, for our assumed inherited initial \ce{CH4}/CO=0.02, C/O cannot rise significantly above 1. Somewhat greater C/O is seen by \citet{Mah_2023}, as their models use a much higher initial \ce{CH4}/CO=0.5, however (as also shown in Fig.\,8 of \citealt{Long_2025}), C/O still rarely exceeds $\sim\!1.2$. This is comfortably below the inner disc C/O$\gtrsim\!2$ inferred for several VLMSs \citep{Long_2025,Kanwar_2025} or DoAr\,33 \citep{Colmenares_2024}. We note that \ce{CH4}/CO would be further lowered if most of the \ce{CH4} were trapped in \ce{H2O} ices such that it only desorbed at higher temperatures \citep[e.g.][]{Collings_2004}.

We conclude that the destruction of \ce{CO}, along with a considerable additional source of hydrocarbons, would be required to reach C/O>1. We next explore how to achieve such conditions.

\subsection{CO transformation}
As the dominant molecule at late times in the models above is CO, which imposes C/O=1, it is imperative that it is depleted in favour of other species if C/O>1 is to be reached.
Reactions alone only reportion a fixed C and O budget between different molecules, but cannot alter the (total) C/O ratio. Instead, C and O must be separated into different molecules with the C remaining in a volatile species and the O becoming sequestered in a less volatile species (e.g. \ce{H2O}) and transported more rapidly out of the disc by radial drift. In the gas phase, this can be achieved by the formation of \ce{CH4}. Where CO is comfortably below its freeze-out temperature, it is predominantly destroyed by hydrogenation to \ce{CH3OH}, which if followed by UV photodissociation, may potentially also lead to \ce{CH4}.
In Appendix \ref{appendix:reactions} we clarify the role played by the different reactions in our CO destruction scheme by adding them one by one for an ionization rate $\zeta_{\rm \ce{H2}}=1.3\times10^{-17}\,\mathrm{s^{-1}}$. We demonstrate that the most crucial reaction is the gas phase route as it directly liberates C. \ce{CH3OH} formation alone does not liberate C and cannot produce C/O>1, and in smooth discs, the timescale to photodissociate the \ce{CH3OH} ice is too short compared to the drift timescale to rectify this. 

Here, we focus on the results of a grid of models including all CO transformation routes. As the models without reactions involving C showed a much stronger dependence on the initial disc size than the initial disc-to-star mass ratio, in Fig. \ref{fig:COdepletion} we only show models with a fiducial $M_{\rm D,0}/M_*=0.05$ but note that models with other mass ratios show very similar behaviours, with a tendency towards slightly faster chemical evolution for lower mass discs as they attenuate CRs slightly less.

For a given set of properties ($M_{\rm D,0}/M_*$, $R_{\rm C,0}$, $\zeta_{\rm \ce{H2}}$), the evolutionary pathways of discs around 1.0 and 0.1\,$M_\odot$ stars are similar (Fig. \ref{fig:COdepletion}).
As C/O>1 is reached by destroying CO, such that the O ends up in less-volatile molecules with a shorter lifetime in the disc, it happens primarily through mild O depletion at roughly solar or slightly enhanced C/H. The maximum C/O occurs when the gas is slightly depleted in C, but very depleted in O.
As seen when reactions are neglected, the evolution is more rapid around the VLMSs as predicted by \citet{Mah_2023}. Thus, as before, by the arbitrary end time of 10\,Myr, VLMSs show significant depletion of both C/H and O/H, while the TTSs end with somewhat depleted O/H, but still relatively high C/H, and, consequently often a higher C/O.

\begin{SCfigure*}[][!ht]
\begin{wide}
    \centering
    \includegraphics[width=0.75\textwidth]{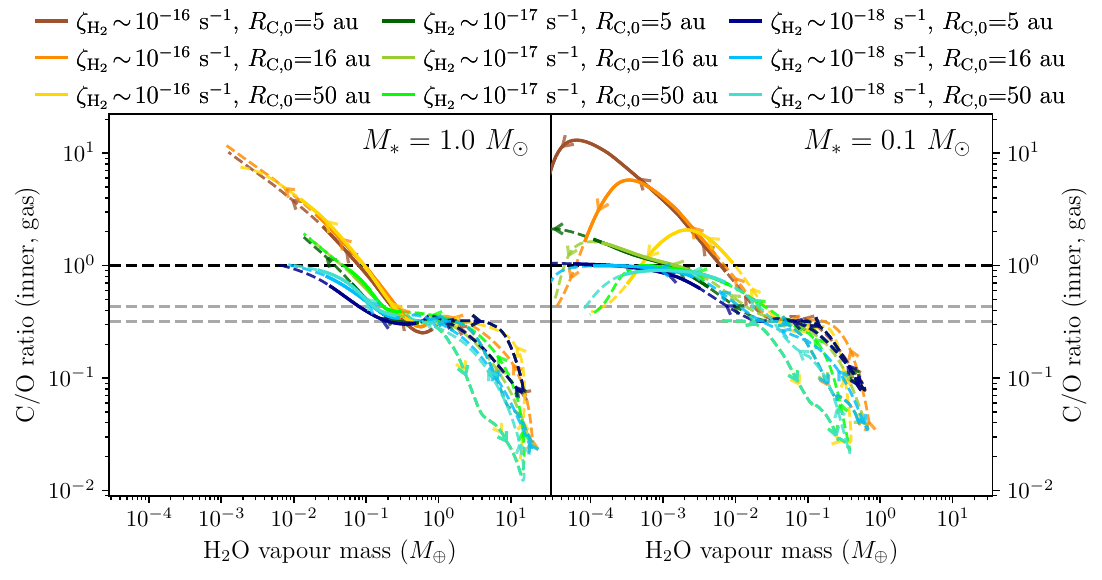}
    \caption{C/O ratio as a function of inner disc \ce{H2O} vapour mass (calculated as the total integrated \ce{H2O} vapour mass inside of the midplane snowline) for the same models as Fig. \ref{fig:COdepletion}. The line styles are as that figure and the arrows illustrate the evolution with time.}
    \label{fig:water}
\end{wide}
\end{SCfigure*}

\subsubsection{Dependence on ionization rate}
The clearest trend in Fig. \ref{fig:COdepletion} is that higher C/O is achieved for higher ionization rates.
Essentially all models for both stellar masses remain under C/O=1 when the ionization rate is decreased to $\sim\!10^{-18}\,\mathrm{s^{-1}}$.
The fiducial ionisation rate is able to create a maximum $1 \lesssim\!\mathrm{C/O} \lesssim\!2$. However, a highly elevated ionization rate of $\sim\!10^{-16}\,\mathrm{s^{-1}}$ can reach $2 \lesssim\!\mathrm{C/O} \lesssim\!10$, consistent with some of the more extreme cases inferred from observations.

This trend can be understood by considering the various timescales at play, relative to the 10\,Myr for which we run our models (only a few rare exceptions, for example J0446B as studied by \citet{Long_2025}, live longer than this).
The average timescale that a molecule m spends in the gas phase before accreting onto the star after being released at its snowline can be estimated, following Eq. 18 of \citet{Sellek_2025}, as
\begin{equation}
    t_{\rm acc,m} = 0.28\,\mathrm{Myr}  \left(\frac{M_*}{M_{\odot}}\right)^{3/2} \left(\frac{\alpha}{10^{-3}}\right)^{-1} \left(\frac{h_0}{0.021}\right)^{2} \left(\frac{T_{\rm snow,m}}{100\,\mathrm{K}}\right)^{-2}
    .
    \label{eq:taccmol}
\end{equation}
Given that we scale $h_0 \propto M_*^{-0.425}$ \citep{Sinclair_2020}, then for our parameters and assuming $T_{\rm snow, CO}=20\,\mathrm{K}$, we find $t_{\rm acc,CO} = 7\,\mathrm{Myr} \left(M_*/M_{\odot}\right)^{0.65}$, i.e. 1.6\,Myr and 7\,Myr for $0.1$ and $1.0\,M_{\odot}$ stars respectively.
For comparison, the timescale for the most critical reaction, \ce{CO ->[He+] CH4}, is $\sim\!4\,\mathrm{Myr}\left(\zeta_{\rm \ce{H2}}/10^{-17}\,\mathrm{s^{-1}}\right)^{-1}$ assuming CO dominates the consumption of \ce{He+}.

Thus, if the ionization rate is lowered, the reaction rate becomes longer than both the 10\,Myr for which we run the models, and the advection timescales, so most CO molecules reach the star without a chance of being converted to \ce{CH4} and maintain a C/O$\sim\!1$.
Conversely if the ionization rate is raised, CO molecules can react in the gas phase in <1\,Myr, much faster than it would take to accrete onto the star around stars of any mass. Consequently, the destruction of CO and formation of \ce{CH4} is efficient in these high $\zeta_{\rm H_2}$ models, resulting in a high C/O within 2-3 Myr for any $M_*$. Only for the fiducial ionization rate of $\sim\!10^{-17}\,\mathrm{s^{-1}}$ do we find that, within the timescale of most observations so far, C/O<1 for the TTSs and >1 for the VLMSs.

\subsubsection{Dependence on disc size}
\label{sec:results_size}
The second trend in Fig. \ref{fig:COdepletion} is a dependence on initial disc size, which becomes especially clear at high ionization rate. Intriguingly, initially smaller discs around VLMSs reach higher C/O, while for the TTSs, initially larger discs are more C-rich for much of their early evolution.
The sensitivity to this parameter is such that at the fiducial ionization rate, whether discs around VLMSs cross the C/O=1 boundary depends on their initial size. Larger ($\gtrsim\!50\,\mathrm{au}$) discs around VLMSs do not reach C/O>1 without an elevated ionization rate.

The short accretion timescales at the CO snowline for the VLMSs mean that they are observed over several such timescales. 
Thus, their behaviour is controlled by how quickly the reservoir of C-rich gas is able to be depleted by accretion to a level where sublimation from the remnant small dust grains comes to dominate and resets C/O to $\sim\!0.32$. The larger the gas-phase reservoir of CO relative to the abundance of small grains, the later this happens and the more time there is to convert CO to \ce{CH4} and raise C/O. Smaller discs have a) a larger fraction of CO (94\% for 5\,au discs compared to 24\% for 50\,au discs) initially in the gas phase (preventing it being instead converted to \ce{CH3OH}), and b) a shorter radial drift timescale and therefore less dust left at late times, and thus reach higher, later peaks in C/O.

Conversely, the TTSs are assumed to have denser discs, leading to more significant attenuation of ionizing particles. In larger discs, the mass is more spread out, making the disc optically thinner. Consequently, a larger fraction of the disc CO gas (46\% for 50\,au discs compared to 8\% for 5\,au discs) experiences ionization rates sufficiently above the $\sim\!10^{-18}\,\mathrm{s^{-1}}$ floor (which alone is insufficient to destroy CO within a disc lifetime).

\subsubsection{Relationship with the presence of water}
As outlined above, small amounts of O-rich ices sublimating from remnant dust can contribute to limiting the growth of C/O.
Figure \ref{fig:water} shows the C/O ratio as a function of the total integrated \ce{H2O} vapour mass in the inner disc.
This reduces the spread between the tracks for different initial radii, demonstrating the evolution of the disc's solids is important even though the critical destruction of CO happens in the gas phase.
The higher the ionization rate, the sooner CO is broken up, and the sooner the O (in the form of \ce{H2O}) can be removed by the action of drift. Thus we see that while fiducial ionization rates make \ce{H2O} slightly more abundant at 10\,Myr as they produce some \ce{H2O} from CO, shifting the tracks to the right compared to the lowest ionization rates, the highest ionization rates result in much lower \ce{H2O} after 10\,Myr.

We note, however, that this may not imply an anti-correlation between C/O and the observable \ce{H2O} emission. The observable \ce{H2O} emission is limited by photodissociation in the UV-exposed upper layers, and the dust continuum optical depth, where the amount of dust may be correlated with the amount of \ce{H2O} delivered to the inner disc resulting in little sensitivity of the column densities to the \ce{H2O} abundance \citep{Sellek_2025,Houge_2025}. Indeed, \ce{H2O} is detected in a few VLMS discs, and at line-to-continuum ratios similar to TTS discs, but is harder to identify due to the strong \ce{C2H2} emission \citep{Arabhavi_2025b}.

\begin{SCfigure*}[][t]
\begin{wide}
    \centering
    \includegraphics[width=0.75\textwidth]{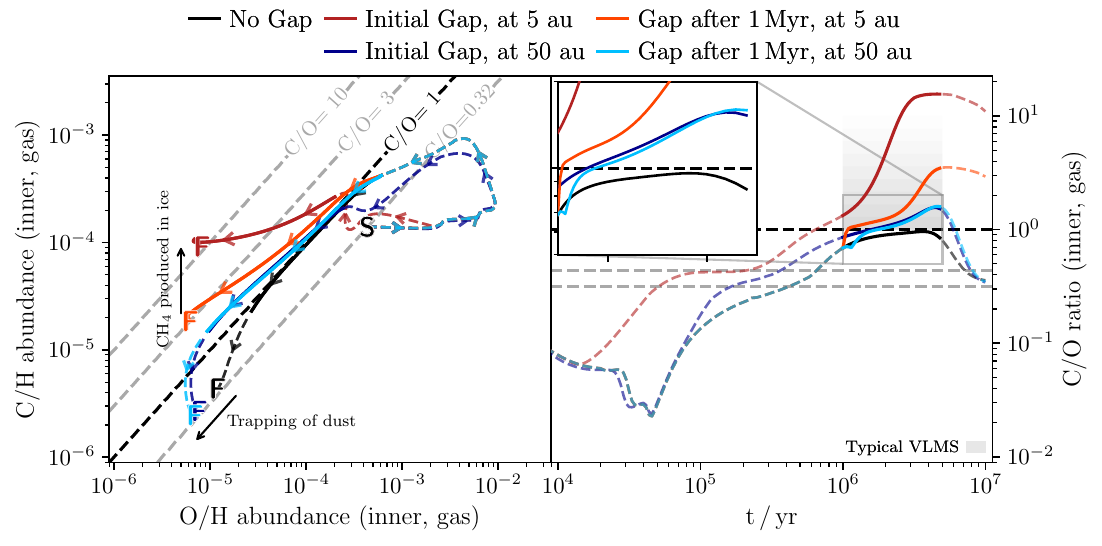}
    \caption{As for the right-hand panels of Fig. \ref{fig:COdepletion} but comparing models with and without dust traps for a $R_{\rm C,0}=50$\,au disc around a $0.1\,M_\odot$ star, with an ionization rate $\sim\!10^{-17}\,\mathrm{s^{-1}}$.}
    \label{fig:trapParams}
\end{wide}
\end{SCfigure*}

\subsubsection{Formation of larger hydrocarbons}
\ce{CH4} has rarely been detected so far in MIR spectra of discs around stars $>0.3\,M_{\sun}$, with tentative exceptions of DoAr33 \citep{Colmenares_2024} and CY Tau \citep{Temmink_2025}. Rather, one of the most commonly detected species in MIR spectra of protoplanetary discs is \ce{C2H2}. VLMS disc spectra do show \ce{CH4} \citep{Arabhavi_2024,Kanwar_2024b}, but also show larger hydrocarbons including \ce{C4H2}, \ce{C6H6}, \ce{C3H4}, \ce{C2H6} and \ce{C2H4} \citep[in order of detection frequency,][]{Arabhavi_2025b}. 
Thus, for completeness, we henceforth use models in which the hydrocarbon formation progresses further to \ce{C2H2}, and beyond, via successive hydrogenation, to \ce{C2H4} and \ce{C2H6}.

These species are all less volatile than \ce{CH4}, but more so than \ce{CO2}.
These less-volatile hydrocarbons are transported inwards by drifting ices to a closer snowline, shortening the lifetime of C-rich gas per Eq. \ref{eq:taccmol} and limiting the maximum C/O somewhat.
Nevertheless, we find that the picture of C/O evolution remains broadly similar to models which only include CO transformation, so we defer a more detailed discussion to Appendix \ref{appendix:hydrocarbons}. 

\subsection{The role of dust traps}
\label{sec:results_traps}
Dust traps have been connected to the depletion of C \citep[e.g.][]{McClure_2020,Sturm_2022} - as well as refractory elements such as Ca \citep{Micolta_2023,Micolta_2024} - in accreting material. Moreover, some works \citep{Banzatti_2020,Banzatti_2023,RomeroMirza_2024b} relate the strength of MIR \ce{H2O} emission to dust disc size - as a proxy for substructures at large $R$ which hold back the dust - although other works \citep{Gasman_2025,Temmink_2025} find less clear trends or distinctions. Thus, we build on other modelling works \citep{Kalyaan_2021,Kalyaan_2023,Easterwood_2024,Mah_2024,Sellek_2025} and explore the impact of dust traps in the outer disc, which reduce the inward pebble flux and subsequent delivery of volatiles to the inner disc.
We consider a model grid (using the hydrocarbon formation scheme) with four options for pressure bumps that trap dust: forming either early or late and at warm or cold temperatures (Table \ref{tab:parameters}). Appendix \ref{appendix:traps} shows the results for all models. The results are similar for $1.0\,M_{\odot}$ stars and all $R_{\rm C,0}$, so here we focus on VLMS models with $R_{\rm C,0}=50\,\mathrm{au}$, which is a) the only size where significant mass lies outside both of the trap locations considered and b) the least effective size at raising C/O in smooth discs, to see if traps can remedy this.

Figure \ref{fig:trapParams} shows how, at a given time, any of the models with traps have a lower O/H than the corresponding model with no trap. The trap must therefore reduce the pebble flux to the inner disc, and thus the supply of \ce{H2O} (a major O-bearing molecule, especially in the presence of CO destruction, e.g. Fig. \ref{fig:water}) which would be frozen out at both traps. Even the cold gap at 50\,au, despite only being able to trap a small fraction of the dust mass, increases C/O to above 1. This trapping results in a similar flux of dust to the inner disc as in the more compact models without traps (which had simply lost their dust more quickly) and consequently the evolution of the C/O is very similar.

Conversely, warm dust traps at 5\,au show a distinct behaviour, where they have significantly higher C/H than a model with no traps, despite a lower O/H; consequently they also sustain a high C/O.
At the location of these traps, \ce{H2O}, \ce{CH3OH}, \ce{CO2} and \ce{C2H_x} hydrocarbons will be frozen out and trapped, while \ce{CH4} and \ce{CO} are in the gas phase and can pass through normally. As the O/H is lowered, the ices cannot be leaking through significantly, yet the trapped material must be the source of additional C.
Instead, the trap results in \ce{CH3OH} ice residing long enough in the disc (unlike in smooth discs, Appendix \ref{appendix:reactions}) to be photodissociated to \ce{CH4} (by the secondary UV created by CRs)\footnote{\citep{Ligterink_2024} also suggested that warm, UV-irradiated dust traps could trap ices long enough for the formation of organic macromolecules.}. This then sublimates and passes through.
We confirmed this by running extra models in which this reaction was switched off, and in which the sustained high C/H was no longer seen.
Warm traps that form early enough to trap a significant mass of \ce{CH3OH} before most of the pebbles drift past 5\,au, can reach C/O>10, while later trap formation leads to lower C/O$\sim\!3$, though these values are upper limits due to the uncertain branching ratios for \ce{CH3OH} photodissociation \citep{Oberg_2009d}.
This scenario involves a large change in the total (gas+ice) C/O across the dust trap, from $\gtrsim\!10$ inside, to $<\!1$ outside (see Fig. \ref{fig:COvsR}). This loosely resembles the model proposed for J1605 by \citet{Kanwar_2025}, where strong O depletion is needed in the hottest gas, but C/O falls again further out (in order to reproduce the \ce{CO2} fluxes), with the two reservoirs separated by a gas gap.
Cold traps do not produce the same effect: any \ce{CH4} produced remains frozen out.

\section{Discussion}
\label{sec:discussion}
\subsection{What is needed for higher C/O around VLMSs?}

\begin{figure}[!ht]
    \centering
    \includegraphics{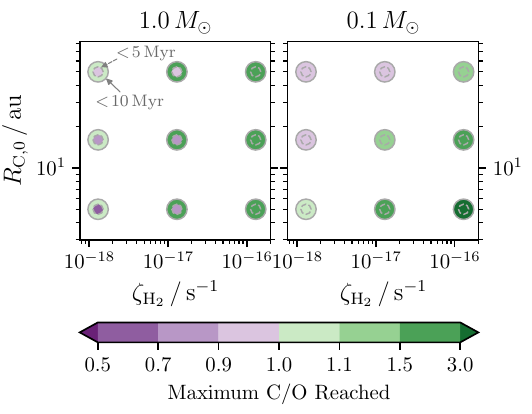}
    \vspace{-12pt}
    \caption{Maximum inner-disc gas-phase C/O achieved for different parameters in smooth disc models with hydrocarbon formation around $1.0\,M_{\odot}$ TTSs (left) or $0.1\,M_{\odot}$ VLMSs (right). Darker green colours indicate more C-rich discs while darker purple colours indicate more O-rich discs; note that the colour bar is neither linear nor symmetric. The central circles show the maxima within the first 5\,Myr, while the outer rings show the maxima across the full 10\,Myr model run.}
    \label{fig:maxCOsummary}
\end{figure}

\begin{figure*}
    \centering
        \includegraphics[width=0.385\linewidth]{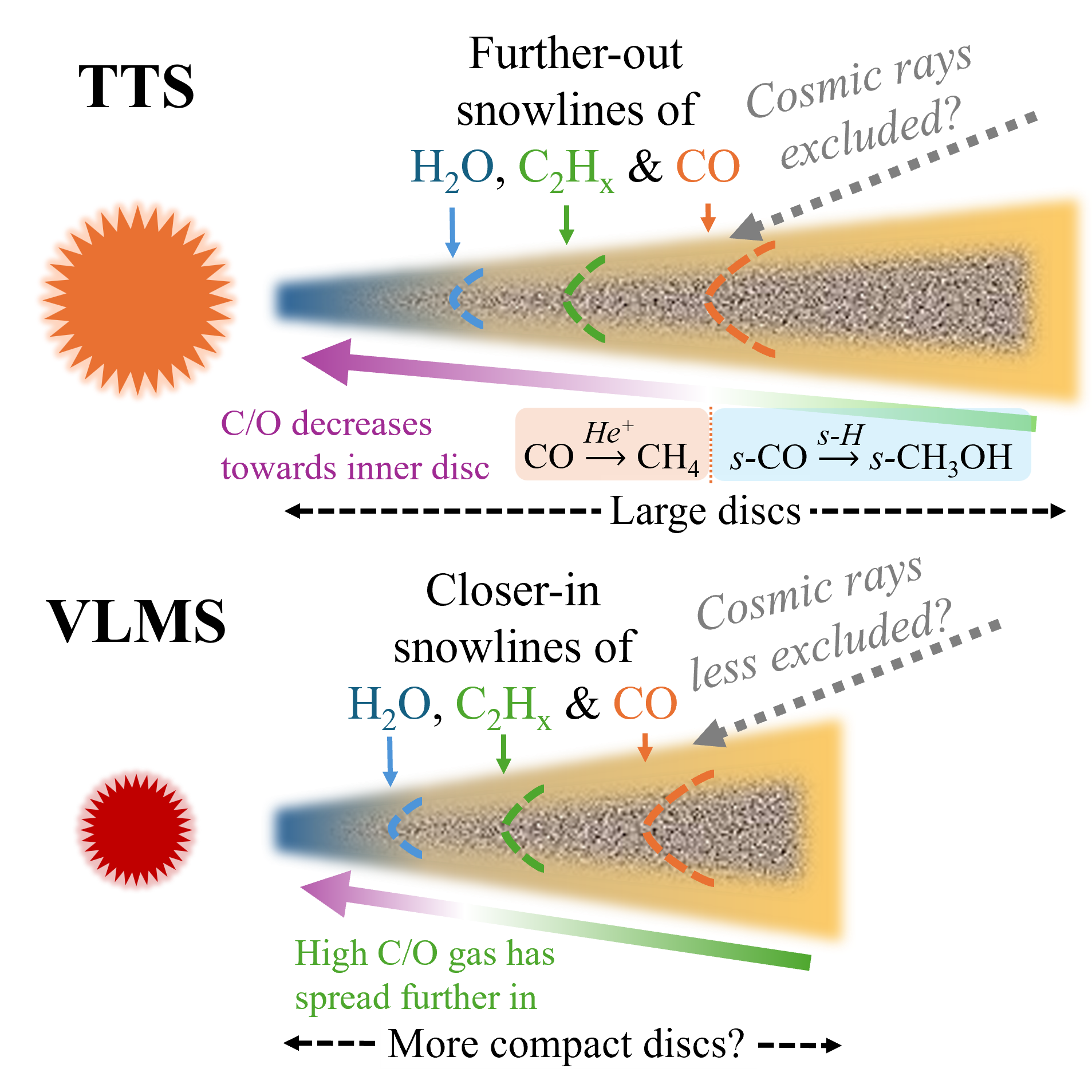}
        \includegraphics[width=0.49\linewidth]{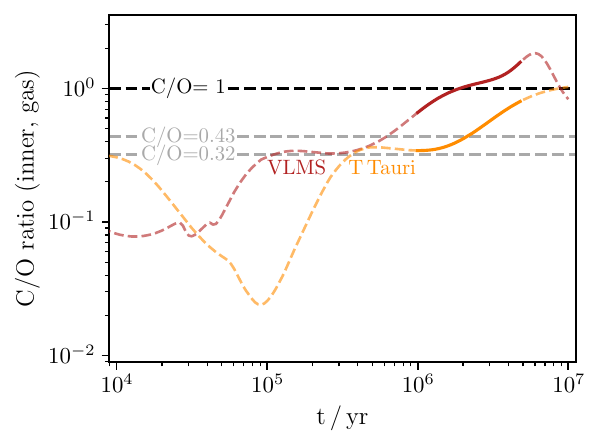}
    \vspace{-8pt}
    \caption{Summary of differences between VLMS discs and TTS discs in this work. For the purposes of this comparison, we assume a TTS model with the typically measured ionization rate of $\sim\!10^{-18}\,\mathrm{s^{-1}}$ and an average $R_{\rm C,0}=16\,\mathrm{au}$. We assume that the VLMS has a more compact disc of $R_{\rm C,0}=5\,\mathrm{au}$ subject to an unexcluded galactic CR ionization rate of $\sim\!10^{-17}\,\mathrm{s^{-1}}$.}
    \label{fig:summary}
\end{figure*}

Figure \ref{fig:maxCOsummary} summarises the maximum C/O achieved in the first 5\,Myr and the full 10\,Myr in the smooth disc models with hydrocarbon formation. 
Our models suggest that for VLMSs, CO transformation combined with dust and gas transport reliably produces C/O>1, as observed for most of their discs, for $\zeta_{\rm H_2}\gtrsim\!10^{-17}\,\mathrm{s^{-1}}$ and $R_{\rm C,0}\lesssim\!16\,\mathrm{au}$. Conversely, for more massive TTSs where C/O is usually <1, our models match this for the first 5\,Myr if $\zeta_{\rm H_2}\lesssim\!10^{-17}\,\mathrm{s^{-1}}$ - though $\zeta_{\rm H_2} \lesssim\! 10^{-18}\,\mathrm{s^{-1}}$ is required to keep discs below C/O=1 for the full 10\,Myr - with weaker dependence on initial disc size.
Figure \ref{fig:summary} thus summarises the differences hypothesised for discs around VLMSs, compared to the typical values measured for TTSs, that would promote a higher C/O: a colder disc with close in snowlines, a more compact disc, and a higher ionization rate (possibly due to weaker exclusion of CRs).
We now discuss the viability of some of these.

\subsubsection{High ionization rates}
\label{sec:discussion_ionisation}
A key ingredient in our model is ionization, which must be at a rate $\gtrsim\!10^{-17}\,\mathrm{s^{-1}}$ for the VLMS discs to become C-rich and $\lesssim\!10^{-17}\,\mathrm{s^{-1}}$ for TTSs to remain O-rich for the first few Myr. ALMA observations have constrained ionization in protoplanetary discs (around $M_*\sim\!0.5-2.1\,M_*$ stars) using the molecular ions \ce{HCO+}, \ce{N2H+}, and their isotopologues \citep{Cleeves_2015,Seifert_2021,Aikawa_2021,Long_2024,Kashyap_2024}. 
These efforts have produced values ranging from $10^{-20}-10^{-18}\,\mathrm{s^{-1}}$, which are compared with the rates assumed in our $1\,M_\odot$ models in Fig. \ref{fig:zetaCR}. 
While some works quote midplane rates and others incident rates, the observations are mostly sensitive to scales of 10s au where, with the 'standard' attenuation of CRs, the ionization would not be reduced at the midplane compared to the incident rate.

As discussed in Section \ref{sec:methods_ionisation}, there are several contributions to the ionization in discs, including galactic CRs (which have the highest energy so most easily penetrate to the midplane), SLRs, and energetic particles (protons) and radiation (X-rays) produced by stellar magnetic activity.
The low ionization rates $\lesssim\!10^{-18}\,\mathrm{s^{-1}}$ inferred by the aforementioned works suggest that the galactic CR flux is reduced. CRs may be more strongly attenuated than assumed e.g. due to non-vertical propagation \citep{Fujii_2022,Long_2024}. 
Alternatively, the incident rate may be severely reduced \citep{Cleeves_2015} due to the exclusion of galactic CRs by the strong, highly magnetic stellar winds of young, rapidly rotating stars \citep{Cleeves_2013}. 
However, while the observed rates meet our $\lesssim\!10^{-17}\,\mathrm{s^{-1}}$ criterion to avoid C-rich TTSs, they would be insufficient to drive the chemistry to C/O>1 around VLMSs. 
This suggests that for their discs to undergo sufficient chemical evolution, VLMSs must have higher ionization rates, either due to negligible exclusion of CRs or a significant additional source of ionization that compensates.

X-rays could be one such additional ionization source. At the ages of interest, most stars $\lesssim\!1.0\,M_\odot$ are still on a Haysashi track and thus are fully convective \citep[e.g.][]{Sokal_2020,Getman_2021}. They are also relatively rapid rotators. 
Thus, both TTSs and VLMSs are highly magnetically active and produce strong X-rays. Both lie in the `saturated regime' where $L_{\rm X}/L_{\rm bol} \sim\!10^{-3}$, \citep{Preibisch_2005}, such that $L_{\rm X}$ increases steeply with $M_*$ \citep[see also][]{Gudel_2007a,Flaischlen_2021}. Assuming a passively heated irradiated disc, material at a given temperature - e.g. at a given snowline - is heated by a given stellar flux ($L_{\rm bol}/r^2$). Given similar $L_{\rm X}/L_{\rm bol}$, the incident X-ray fluxes ($L_{\rm X}/r^2$) should thus also be similar at a given snowline\footnote{This is different to the main sequence, where M dwarfs have a higher level of magnetic activity than solar-type stars due to their faster rotation periods and deeper convective zone, resulting in a higher $L_{\rm X}/L_{\rm bol}$. Then, orbiting planets at a given temperature, for example in the habitable zone, would experience a higher X-ray flux \citep{Johnstone_2021}.}. While flares are common and strong, their time-averaged luminosity is of the order of $\sim\!10\%$ of the background `characteristic' level for all spectral types \citep{Audard_2000,Getman_2021,doAmaral_2022} and so are unlikely to alter this trend. Moreover, X-rays seem unlikely to dominate ionization or drive chemistry at the midplane \citep{Cleeves_2013,Cleeves_2014b,Rab_2017}. This is distinct from the disc surface, where VLMSs have a higher $L_{\rm X}/L_{\rm UV}$ that changes the dominant chemical pathways \citep[][we note that their M dwarf model used a roughly maximal $L_{\rm X}$ for a $0.1\,M_\odot$ star]{Walsh_2015}. Nevertheless, the X-ray-dominated layer may penetrate closer to the midplane in lower mass discs.

There are next to no direct constraints on stellar energetic particles, though it has been suggested that they could explain potentially higher ionization rates close to DM Tau \citep{Long_2024}.
Young stars, along with other active stars and individual flares, follow a G{\"u}del-Benz relation \citep{Gudel_1993,Benz_1994}: a correlation between their X-ray and gyrosynchotron luminosities \citep{Dzib_2013,Dzib_2015,Pech_2016}.
Gyrosynchotron radiation traces the acceleration of relativistic electrons by magnetic activity, with protons, which dominate the stellar energetic particle flux, likely also accelerated.
This correlation could tentatively suggest that VLMSs, which have lower $L_{\rm X}$, would produce fewer energetic particles than earlier type stars, thus going against the trend we would need. 

Finally, both solar-type and late M-dwarf stars are expected to have stellar winds that can exclude CRs \citep[e.g.][]{RodgersLee_2020b,Mesquita_2022}. The CR energies that are excluded at a distance $r$ are those for which the advective timescale in the wind $t_{\rm adv} = r/v_{\rm wind}$ is shorter than the diffusive timescale $t_{\rm diff}=r^2/\kappa$, where the diffusivity $\kappa \propto 1/B$, the magnetic field strength. 
Thus, if VLMSs launch slower or more weakly magnetic winds, they would exclude CRs less effectively. 
$B$ declines at least as fast as $r^{-1}$, making $t_{\rm adv}/t_{\rm diff}$ an increasing function of radius. Thus, CR exclusion should be weaker further from the star \citep{Langner_2005,Mesquita_2022}, as suggested to explain the increase in IM Lup's ionization rate beyond 100\,au \citep{Seifert_2021}. Therefore, on e.g. $\sim\!10\,\mathrm{au}$ scales, VLMSs may have to make up less than the 3 orders of magnitude difference implied by comparing directly to the $\zeta_{\rm H_2, 1\,au}\sim\!10^{-20}\,\mathrm{s^{-1}}$ that \citet{Cleeves_2013} estimated by extrapolating from the CR fluxes reaching Earth.
Overall, it seems possible that VLMSs could exclude CRs less efficiently than TTSs but modelling of CR propagation as a function of distance around pre-MS stars of different masses is required to confirm if the exclusion is sufficiently negligible to reach $\zeta_{\rm H_2}\sim10^{-17}\,\mathrm{s^{-1}}$ within the CO snowline.

\subsubsection{Small initial disc sizes}
Compact initial discs are needed around VLMSs to allow them to deplete enough dust to reach C/O significantly greater than 1 (unless $\zeta_{\rm H_2}$ is high enough that TTSs would also host carbon-rich discs with 5\,Myr). Thus, a dependence of disc size on stellar mass would help ensure that the VLMSs reach a consistently higher inner disc C/O than the more massive TTSs.

Theory predicts that discs may form more compact around lower mass stars  \citep[e.g.][]{Hennebelle_2016}. This could place them towards the lower end of our adopted range, which was used for both $M_*$ but based on a sample skewed towards higher masses \citep{Trapman_2023}. Observations of embedded protostellar discs may hint in this direction \citep[e.g.][]{Yen_2017} but at the Class II stage, only the few brightest (and hence largest) VLMS discs have been studied so far \citep{Kurtovic_2021}.
More measurements of gas disc sizes around VLMSs are needed to confirm if they are indeed systematically smaller. 

\subsubsection{Warm dust traps}
Section \ref{sec:results_traps} shows that \ce{CH3OH} photodissociation at a dust trap inside the \ce{CH4} snowline could provide a supply of additional C to the inner disc, potentially producing quite extreme C/O in discs that do not otherwise reach C/O>1. The known gaps on $\lesssim\!5\,\mathrm{au}$ scales \citep{GuerraAlvarado_2025} span spectral types K7 to M5 so include both TTS discs and VLMS discs. While this scenario could explain unusual or extreme cases such as DoAr\,33 or J1605 respectively, we consider it an unlikely explanation for the C-rich - but less extreme - chemistry of most VLMS inner discs given that gap opening is not expected to be easier for lower $M_*$ \citep{Sinclair_2020} and substructured discs indeed appear rarer around lower mass stars \citep[in line with the occurrence rates of sufficiently massive exoplanets][]{vanderMarel_2021c}. Thus it is more probable that such a scenario would occur around a TTS, rather than a VLMS.

\subsection{The need for more inner disc C/O measurements}
At present, there are few robust MIR measurements of inner disc C/H, O/H, and C/O; mostly what has been inferred is whether C/O<1 or C/O>1.
More accurate values would help discriminate whether, for example, a `standard' cosmic ray ionization rate around VLMSs of $\zeta_{\rm \ce{H2}, CR}\sim\!10^{-17}\,\mathrm{s^{-1}}$ suffices or whether the C/O is extreme enough that their discs would need to receive enhanced fluxes of energetic particles to explain the observations. Likewise, it would be useful to know how close the discs which do not show C/O>1 come to doing so, and to know elemental abundances in order to understand if chemical transformation is nevertheless at work in some way.
A reliable method must account for effects such as radial gradients in molecular abundances and variations in their respective emitting areas/locations; developing a reliable proxy in terms of retrieved column densities that averages over these effects would be useful.

Moreover, measurements across a wider range of ages would break some degeneracies.
We have argued that a minimum ionization rate of $10^{-17}\,\mathrm{s^{-1}}$ is needed to explain the C-rich inner discs of most VLMSs while a maximum of $10^{-17}\,\mathrm{s^{-1}}$ is needed to explain the O-rich inner discs of most TTSs. It is thus not clear whether we require higher ionization rates around the VLMSs, as $10^{-17}\,\mathrm{s^{-1}}$ could apply to all discs around stars of all masses. However, this value would lead to discs around TTSs becoming C-rich (i.e. reaching C/O>1) on timescales of $\sim\!10\,\mathrm{Myr}$, while lower ionization rates would keep these disc below C/O=1.
Studying discs around TTSs in older star-forming regions such as Upper Scorpius and determining whether they remain below C/O=1 would help distinguish between these two outcomes. JWST Cycle 2 GO programs 2970 (PI: Pascucci) and 3034 (PI: Zhang) should shed light on this matter; finding that the discs are C-rich would suggest that the lower C/O seen so far around TTSs is purely a question of the slower physical evolution of the gas and dust, whereas if the discs remain below C/O=1, that would require a slower chemical evolution driven by weaker ionization.

\subsection{The need for complementary observations with ALMA}
As the models presented here involve the transport of chemically-evolved gas from the outer disc to the inner disc, another way to test the origin of the high C/O could be to directly probe the outer disc chemical evolution around VLMSs and compare to that around TTSs using ALMA.
Figure \ref{fig:innerouter} in Appendix \ref{appendix:outer} shows how our models predict that inward transport would cause the inner disc chemistry to lag behind that of the outer disc and thus we expect C/O\textsubscript{in}<C/O\textsubscript{out}.
So far, however, only a small survey of discs around five M4-M5 stars was conducted by \citet{Pegues_2021}, finding that the disc chemistry was fairly similar to TTSs, though with stronger \ce{C2H} described as reminiscent of the hydrocarbon-rich inner discs. 

More measurements of the chemistry of the outer discs of VLMSs are required to build on those initial results. For example, one could attempt to constrain C/O using CO and \ce{C2H} \citep[e.g.][]{Miotello_2019}, C/H using CO and \ce{N2H+} \citep{Anderson_2022,Trapman_2022} or [\ion{C}{1}] \citep{Sturm_2022}, or the ionization rate using \ce{N2H+} or \ce{HCO+} \citep{Cleeves_2015,Seifert_2021,Aikawa_2021,Long_2024,Kashyap_2024}. These would measure, respectively, whether there is high C/O gas in the outer disc that can be transported to the inner disc, whether (chemical) depletion of CO is operating more or less efficiently around VLMSs, and whether there is sufficient ionization to drive the chemistry needed to destroy CO. At present, there are no measurements of whether the ionization rate is higher for VLMSs, as we suggest is required (Section \ref{sec:discussion_ionisation}).

\section{Conclusions}
\label{sec:conclusions}
Distinct differences have been identified between the chemistry of the inner regions of protoplanetary discs around VLMSs ($\lesssim\!0.3\,M_{\odot}$) and more massive TTSs ($\gtrsim\!0.3\,M_{\odot}$). The former tend to have strong hydrocarbon emission, implying a more C-rich chemistry with gas-phase elemental C/O likely >1, while the latter are more O rich. Various suggestions have been put forward to explain these findings. For example, models for inner disc chemical evolution have previously predicted that the C/O should rise after the first $\sim\!1$\,Myr (within which large amounts of \ce{H2O} are delivered to the inner disc), and do so more rapidly for the colder discs around VLMSs \citep{Mah_2023}. However such models are typically limited to C/O$\sim$\!1 (Fig. \ref{fig:noConv}) by having CO as the dominant C-carrying molecule. In this work we explore such models further by demonstrating that in order to reach C/O substantially greater than 1, the CO must be chemically depleted in such a way that the C is eventually liberated.
The chemical transformation follows pathways commonly discussed in the context of underabundant CO (with respect to ISM levels) in the outer disc \citep[][Fig. \ref{fig:reactionScheme}]{Bosman_2018b,Krijt_2020}; in the cold, dark disc midplane, such pathways are ultimately driven by the presence of ionizing energetic particles and/or radiation.

\begin{enumerate}
        \item The most critical reaction to allow C/O>1 is the dissociative ionization of CO in the gas phase by \ce{He+}, which directly liberates C, making it available for forming simple hydrocarbons (with O locked up in \ce{H2O} ice). The less volatile the hydrocarbons formed, the shorter the lifetime of C-rich gas in the disc. This initially raises C/O slightly faster, but limits the maximum C/O reached (Figs. \ref{fig:COdepletion}, \ref{fig:hydrocarbons}).
        \item Whether a disc becomes C-rich, and when, depends on the relative physical and chemical timescales (Fig \ref{fig:maxCOsummary}). In a disc subject to a low ionization rate $\lesssim\!10^{-18}\,\mathrm{s^{-1}}$, the chemical timescales are longer than typical disc lifetimes and C/O remains $\lesssim\!$1. For a very high ionization rate $\gtrsim\!10^{-16}\,\mathrm{s^{-1}}$, the conversion of CO to hydrocarbons and the corresponding O depletion of the gas happens much more quickly than the timescale on which gas moves inwards through the disc, such that any disc would reach C/O>1 within $1-3$\,Myr. 
        \item While, due to shorter physical timescales for their gas transport, discs around VLMSs increase in C/O more quickly and reach their maximum C/O sooner than those around TTSs, they nevertheless need $\zeta_{\rm H_2}\gtrsim\!10^{-17}\,\mathrm{s^{-1}}$ to transform enough CO into hydrocarbons to reach C/O>1. Given that observations suggest $\zeta_{\rm H_2}\lesssim\!10^{-18}\,\mathrm{s^{-1}}$ around stars $0.5-2.0\,M_\odot$ (Fig. \ref{fig:zetaCR}), inline with needing to be $\lesssim\!10^{-17}\,\mathrm{s^{-1}}$ to remain at C/O< for TTSs, VLMSs likely need to experience higher ionization rates, and thus faster chemical conversion, potentially requiring less-effective exclusion of CRs.
        \item To reach high C/O, it is crucial to ensure a very low  resupply of \ce{H2O} vapour at late times (Fig. \ref{fig:water}). Discs around VLMSs therefore need to be at least as compact as the average TTS disc ($\lesssim16\,\mathrm{au}$)- as discs that are initially more compact lose their dust more quickly - or have effective trapping of dust and ices in pressure bumps, in order to reach C/O>1.
        \item The presence of a warm dust trap between the \ce{CH4} and \ce{CH3OH} snowlines allows the \ce{CH3OH} ice that formed outside the CO snowline to remain in the disc for long enough to be photodissociated by the secondary UV field produced by CRs (Fig. \ref{fig:trapParams}). Assuming that \ce{CH4} is the primary product of this process, it can sublimate and escape through the gap, leading to a sharp C/O gradient with a sustained high C/O inside the gap compared to outside (Fig. \ref{fig:COvsR}). This could potentially explain discs with more extreme C/O such as J1605. 
\end{enumerate}

We conclude that CO destruction and the subsequent transport of O-depleted material from the outer disc is a viable way to produce the higher C/O seen in the inner disc around VLMSs, provided that the ionization rates are $\gtrsim10^{-17}\,\mathrm{s^{-1}}$.
Constraints on inner disc chemistry for older discs should demonstrate whether discs around TTSs also reach these conditions - and the differences are just a matter of their longer physical timescales - or whether the chemical transformation is also faster in the VLMS discs. More observations of VLMSs with ALMA could help confirm whether they do have suitable disc sizes, ionization rates, and chemistries for our proposed route to C/O>1.

\begin{acknowledgements}
    We thank the anonymous reviewer for helpful suggestions which improved the clarity of the work.
    We also thank L. Trapman and A. Bosman for sharing details of the CO chemical network, C. Walsh for discussions about the grain surface processes and the role of X-rays in chemistry, J. Kanwar \& I. Kamp for insights about determination of C/O from spectra, N. Kurtovic for discussions about ALMA observations, and D. Rodgers-Lee \& L. I. Cleeves for discussions about ionization processes.
    A.D.S. and E.v.D. acknowledge support from the ERC grant 101019751 MOLDISK.
    E.v.D. also acknowledges support from the Danish National Research Foundation through the Center of Excellence “InterCat” (DNRF150) and grant TOP-1 614.001.751 from the Dutch Research Council (NWO).
\end{acknowledgements}

%
%

\bibliographystyle{aa}
\bibliography{biblio}

\begin{appendix}

\section{Additional details of chemical model}
\label{appendix:chemistry}
\subsection{Molecular data and tunnelling barriers}

\vspace{-12pt}
\begin{table}[h]
    \centering
    \caption{Binding energies, prefactors, and initial abundances (with respect to H) assumed for the species considered in this work and the resulting snowlines around 0.1 and 1.0 $M_\odot$ stars (defined as the locations where the gas and ice abundances are equal at $t=0$).}
    \setlength{\tabcolsep}{3pt}
    \begin{tabular}{cccccc}
    \hline \hline
        Species & $\nu\,/\,\mathrm{s^{-1}}$  &   $E_{\rm bind}$\,/\,K & Abundance & Snowline\,/\,au\\
    \hline
         \ce{H}\tablefootmark{a}    & $1.54\times10^{11}$   &  450  & - &  \\
         \ce{OH}\tablefootmark{a}   & $3.76\times10^{15}$   &  5698 & - &  \\
     \hline
         \ce{CO}                    & $4.1\times10^{13}$    &  910  & $1.2\times10^{-4}$ &   13/25 \\
         \ce{O2}                    & $1.3\times10^{14}$    &  1030 & $0.85\times10^{-4}$ &   12/21 \\
         \ce{CH4}                   & $2.5\times10^{14}$    &  1140 & $2.3\times10^{-6}$ &   8.4/16 \\
         \ce{C2H4}                  & $4\times10^{15}$      &  2200 & - & 2.4/4.4 \\
         \ce{C2H6}                  & $6\times10^{16}$      &  2600 & - & 1.9/3.5 \\
         \ce{C2H2}                  & $3\times10^{16}$      &  2800 & - & 1.6/2.8 \\
         \ce{CO2}                   & $1.1\times10^{15}$    &  2980 & $2.1\times10^{-5}$ &   1.1/2.1 \\
         \ce{CH3OH}                 & $5.0\times10^{14}$    &  4850 & $2.3\times10^{-6}$ &   0.37/0.62 \\
         \ce{H2O}                   & $1.3\times10^{18}$    &  6722 & $1.1\times10^{-4}$ &   0.23/0.45 \\
    \hline
    \end{tabular}
    \newline
    \tablefoottext{a}{Only relevant for grain-surface reactions}
    \label{tab:binding}
\end{table}
\vspace{-12pt}
\begin{table}[h]
    \centering
    \caption{Tunnelling barriers assumed for grain-surface reactions.}
    \begin{tabular}{cc}
    \hline \hline
         Reaction & Tunnelling Barrier\,/\,K \\
     \hline
         \ce{CO + OH \to CO2 + H} & 400 \\
         \ce{CO + H \to HCO} & 2500 \\
         \ce{C2H2 + H \to C2H3} & 1210 \\
         \ce{C2H4 + H \to C2H5} & 750 \\
     \hline
    \end{tabular}
    \label{tab:barriers}
\end{table}
\vspace{-12pt}

\subsection{Equilibrium processes for reactive intermediaries}
For each of the reactive intermediaries that react with CO - \ce{He+}, \ce{H}, and \ce{OH} - we assume equilibrium abundances between a formation rate driven directly by cosmic ray ionisation (CR) or their secondary UV field (CRPHOT) and all major destruction routes, following those in \citep{Krijt_2020}:
\begin{align*}
    \ce{He + CR &-> He+} \\ \ce{He+ + \{CO,N2,H2,dust\} &-> He + \{CO+,N2+,H2+,dust+\}}
    ,
\end{align*}
\vspace{-12pt}
\begin{align*}
    \ce{H2O(s) + CRPHOT &-> OH(s) + H(s)} \\ \ce{OH(s) + \{CO(s),H(s)\} &-> \{CO2(s) + H(s),H2O(s)\}}
    ,
\end{align*}
\vspace{-12pt}
\begin{align*}
    \ce{H2 + CR &->[2.4] H} \\ \ce{H &-> H(s)} \\ \ce{H(s) + H2S(s) &->[3] H2S2(s)}
    .
\end{align*}
Here, \ce{H2S} provides a baseline sink of H atoms on grain surfaces (we do not follow S-bearing species explicitly so assume a constant abundance of $2\times10^{-8}$) and the numbers above the reactions scale the rates of the reactions shown to account for other pathways following \citet{Krijt_2020}.

\subsection{Ionization}
\begin{SCfigure*}[][h]
\begin{wide}
    \centering
    \includegraphics{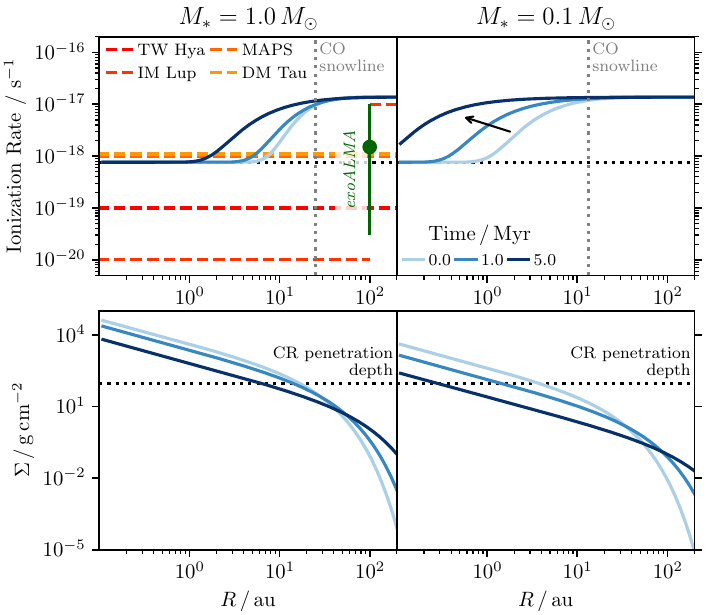}
    \caption{Top: Midplane ionization rate profile at t=0, 1, and 5\,Myr (increasingly dark blues, evolution indicated by arrow) for fiducial models around TTS (left) and VLMSs (right). The TTS models are compared to observationally estimated values for individual sources (red and yellow colours) and the range and median for the exoALMA program \citep{Trapman_2025}. The black horizontal line on each plot represents the floor due to SLRs and the vertical dotted line shows the location of the CO snowline. Bottom: the surface densities of the gas at the times indicated in the top panels. The horizontal line shows the assumed $96\,\mathrm{g\,cm^{-2}}$ penetration depth of CRs.}
    \label{fig:zetaCR}
\end{wide}
\end{SCfigure*}

A key input of our model is the ionization, which creates the intermediaries that reach with CO (and other species). Figure \ref{fig:zetaCR} shows the ionization rate profile at the midplane for our models: for a fixed incident ionization rate at the disc surface ($10^{-17}\,\mathrm{s^{-1}}$ is shown here), the midplane rate varies as a function of time as the disc becomes less optically thick and the attenuation reduces - for comparison, the bottom panel shows the evolving surface densities. Observational estimates \citep{Cleeves_2015,Seifert_2021,Aikawa_2021,Long_2024} are included for the TTSs. In addition, we include the median and range of the ionization rate posteriors derived from CO and \ce{N2H+} fluxes when kinematically derived masses are used as priors, as derived in the exoALMA program by \citep{Trapman_2025}.

\FloatBarrier

\section{Role of different processes}

\subsection{Impact of each molecular reaction}
\label{appendix:reactions}
\begin{figure*}
    \centering
    \includegraphics[width=0.75\linewidth]{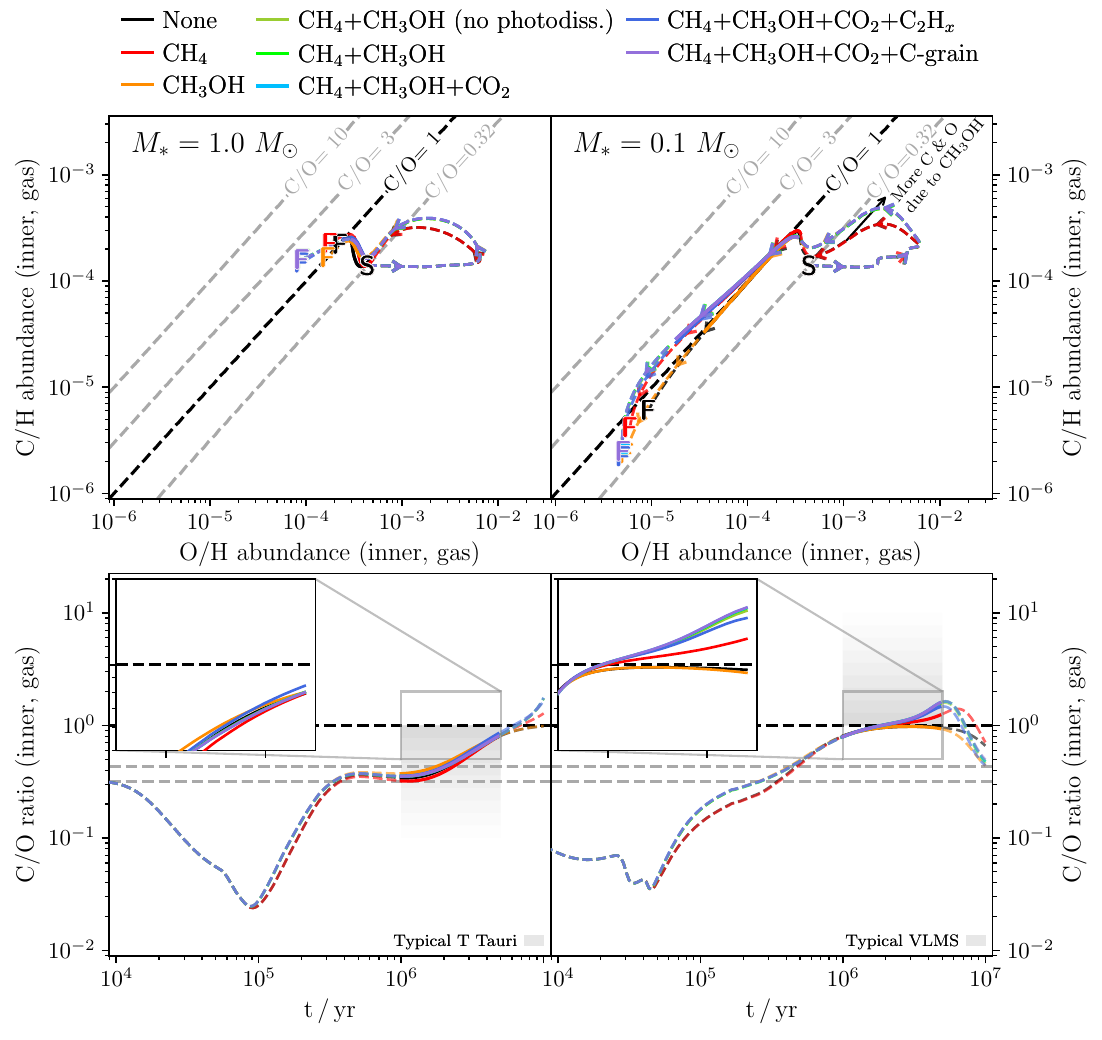}
    \vspace{-8pt}
    \caption{
    As with Fig. \ref{fig:noConv} but for disc models with different numbers/types of reactions included.
    }
    \label{fig:onebyone}
\end{figure*}

To test which reactions matter most, we add them into our model one by one, for a fiducial case with $R_{\rm C}=16\,\mathrm{au}$, $M_{\rm D}/M_*=0.05$, and $\zeta_{\rm \ce{H2}}=1.3\times10^{-17}\,\mathrm{s^{-1}}$.
The results are shown in Fig. \ref{fig:onebyone} and compared to a reference model where no reactions were included and C/O reaches a maximum of 0.98 (black).
The maximum C/O reached in each case is tabulated in Table \ref{tab:onebyone}. For the TTS models, the C/O always rises monotonically over time after the first Myr so a greater maximum is reached after 10\,Myr than within 5\,Myr; the VLMS models typically peak just after 5\,Myr so behave likewise.

The first reaction considered is the conversion of CO to \ce{CH4} (red). We find that this is sufficient for discs around either stellar mass to cross C/O=1 within 10\,Myr, although only the discs around VLMSs cross within the first 5\,Myr. This reaction is key to achieving high C/O levels.

Next we add hydrogenation of CO to \ce{CH3OH} (dark green). While this does not itself separate C and O from one another, it does help flush the most volatile reservoir of O out of the disc more quickly, as shown by the model tracks moving up and to the right in the region where C/O<1. This prevents much of the delivery of CO to its snowline and thus reduces the amount available to buffer the C/O ratio to 1, thus enhancing the effectiveness with which the conversion of gas phase CO to \ce{CH4} raises C/O. 
We also tested a model where this is the only reaction (orange), confirming that while it helps to accelerate the reduction in C/H and O/H, it never leads to C/O>1 on its own. 

The third reaction we add is the photodissociation of the \ce{CH3OH} ice forming \ce{CH4} (light green). While this process can play the crucial role of breaking the CO bond and separating out C and O into species with different volatilities, in practice it makes barely any discernible difference to the evolution of our models, raising C/O by only a percent or two. This is because the timescale for these reactions is on the order of Myr, while the majority of the ice has a much shorter lifetime in smooth discs due to efficient radial drift.

The final route for CO transformation is the reaction with \ce{OH} radicals forming \ce{CO2} (light blue). We find this mostly has no impact on our results as a very limit range of radii at the midplane are within the right temperature range for this to dominate over the other gas or ice phase routes.

In addition, we add three reactions forming the larger hydrocarbons \ce{C2H2}, \ce{C2H4}, and \ce{C2H6} (dark blue). At intermediate times, they slightly accelerate the delivery of C-bearing molecules to the inner disc, raising C/O. Conversely, at late times they have very small impact in the direction of lowering the C/O ratio, as they convert \ce{CH4} into less volatile species, thus reducing the lifetime of the carbonaceous component in the disc and preventing such high C/O from building up.

\begin{table*}[t]
    \centering
    \caption{Maximum C/O achieved within the time range specified for models including different numbers/types of reactions for a fiducial case with $R_{\rm C}=16\,\mathrm{au}$, $M_{\rm D}/M_*=0.05$, and $\zeta_{\rm \ce{H2}}=1.3\times10^{-17}\,\mathrm{s^{-1}}$.}
    \begin{tabular}{ccccc}
    \hline \hline
         & \multicolumn{2}{c}{$1.0\,M_{\odot}$} &  \multicolumn{2}{c}{$0.1\,M_{\odot}$} \\
         Reactions Included  &   t<5\,Myr  &  t<10\,Myr   &   t<5\,Myr  &   t<10\,Myr \\
    \hline
         None       &   0.79	   &  0.98       &   0.98	      &   0.98 \\
         \ce{CH4}   &   0.79	   &  1.28       &   1.23	      &   1.40 \\
         \ce{CH3OH} &   0.80	   &  0.99       &   0.98       &   0.98 \\
         \ce{CH4}+\ce{CH3OH} (no photodiss.)           &   0.80	    &  1.73       &   1.55	   &   1.59 \\
         \ce{CH4}+\ce{CH3OH}           &   0.80	    &  1.74       &   1.58	   &   1.63 \\
         \ce{CH4}+\ce{CH3OH}+\ce{CO2}  &   0.80	    &  1.76       &   1.58	   &   1.63 \\
         \ce{CH4}+\ce{CH3OH}+\ce{CO2}+\ce{C2Hx}  &   0.84	    &  1.65       &   1.46	   &   1.47 \\
         \ce{CH4}+\ce{CH3OH}+\ce{CO2}+C-grain    &   0.80	    &  1.76       &   1.59	   &   1.64 \\
    \hline
    \end{tabular}
    \label{tab:onebyone}
\end{table*}

\subsection{Impact of hydrocarbon formation}

\begin{SCfigure*}[][!t]
\begin{wide}
    \centering
    \includegraphics[width=0.75\textwidth]{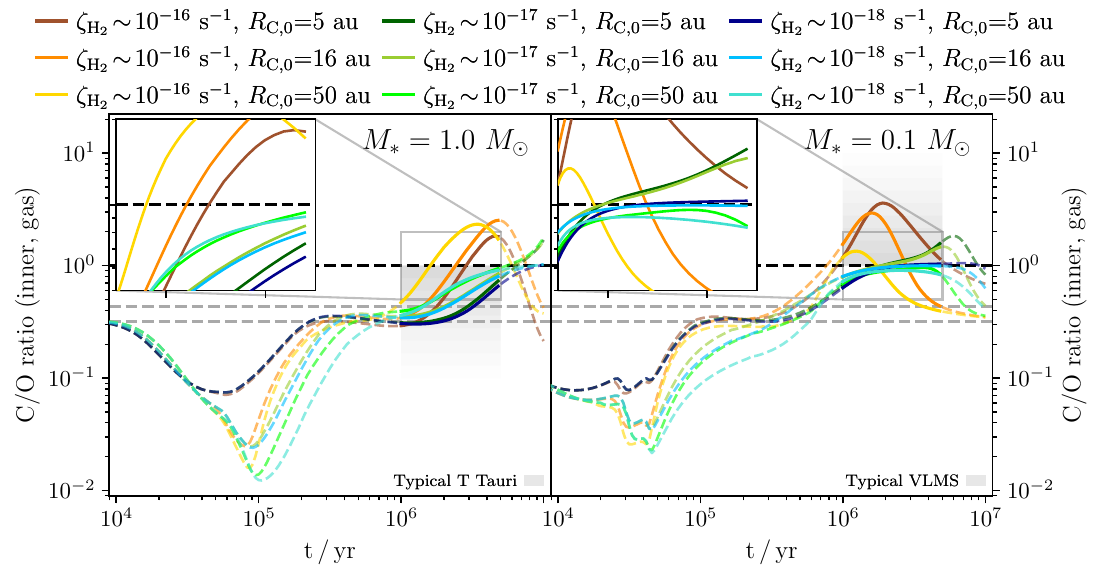}
    \caption{
    As for the lower panels of Fig. \ref{fig:COdepletion} but for disc models which include formation of \ce{C2H_x} hydrocarbons.
    }
    \label{fig:hydrocarbons}
\end{wide}
\end{SCfigure*}

\label{appendix:hydrocarbons}
For ionization rates $\lesssim\!10^{-17}\,\mathrm{s^{-1}}$, the evolution of C/O (likewise C/H and O/H) is very similar when we include larger hydrocarbons (compare Figs. \ref{fig:hydrocarbons} and \ref{fig:COdepletion}). This is because, due to the long chemical timescales, large reservoirs of C \& O remain in CO and are unaffected by the addition of a route out of \ce{CH4}.
The fiducial ionization rates are more visibly affected, but the conclusions remain the same. The timescale for \ce{C2H2} production from \ce{CH4} in our model is $\sim\!3\,\mathrm{Myr}$ - comparable to the accretion timescale at the \ce{CH4} snowline for the TTS model and longer than that for the VLMS model - so there is hardly any time for a \ce{CH4} molecule once formed to be converted to larger hydrocarbons.  

A much larger difference emerges for high ionization rates. Previously, \ce{CH4} was the dominant molecule but now it is efficiently converted to \ce{C2H2}, which is much less volatile. This results in C moving from the gas phase into ices and being transported inwards much faster by drift, such that it is ultimately removed from the disc much sooner. This prevents such a large gaseous C/O ratio from building up, though values $\sim\!3$ are still achieved for both $M_*$.
The formation of less volatile hydrocarbons still allows high ionization rate models to explain DoAr\,33, but they would not be able to reach the extreme levels of O depletion inferred for J1605 by \citet{Kanwar_2025} without also having considerable depletion of C, and thus too low C/O.
We note that trapping of \ce{CH4} in mixed ices would have a similar effect as it would desorb at high temperatures \citep{Collings_2004}, but as much of the \ce{CH4} produced in our models is inside of its snowline, not all \ce{CH4} would be affected by this.

For VLMSs, the endpoint after 10\,Myr is similar, as the gas accretion of \ce{CH4} onto the star around cooler stars (Eq. \ref{eq:taccmol}) is fast enough to have removed it anyway, regardless of conversion into less volatile species.
Conversely, for TTSs, loss of \ce{CH4} via accretion is insufficiently fast, so \ce{C2H2} formation lowers the C/O - and underlying C/H - by over an order of magnitude. 
Such C/H values are comparable to the 1-2 orders of magnitude depletions measured from NIR atomic lines \citep[tracing accreting gas inside the sublimation radius][]{McClure_2019,McClure_2020,Sturm_2022}, but only occur in our models on timescales longer than the ages of most of those sources. 

\subsubsection{Molecular ratios involving \ce{C2H2}}
\label{appendix:molecules}
\begin{figure*}[th]
    \centering
    \includegraphics[width=0.75\linewidth]{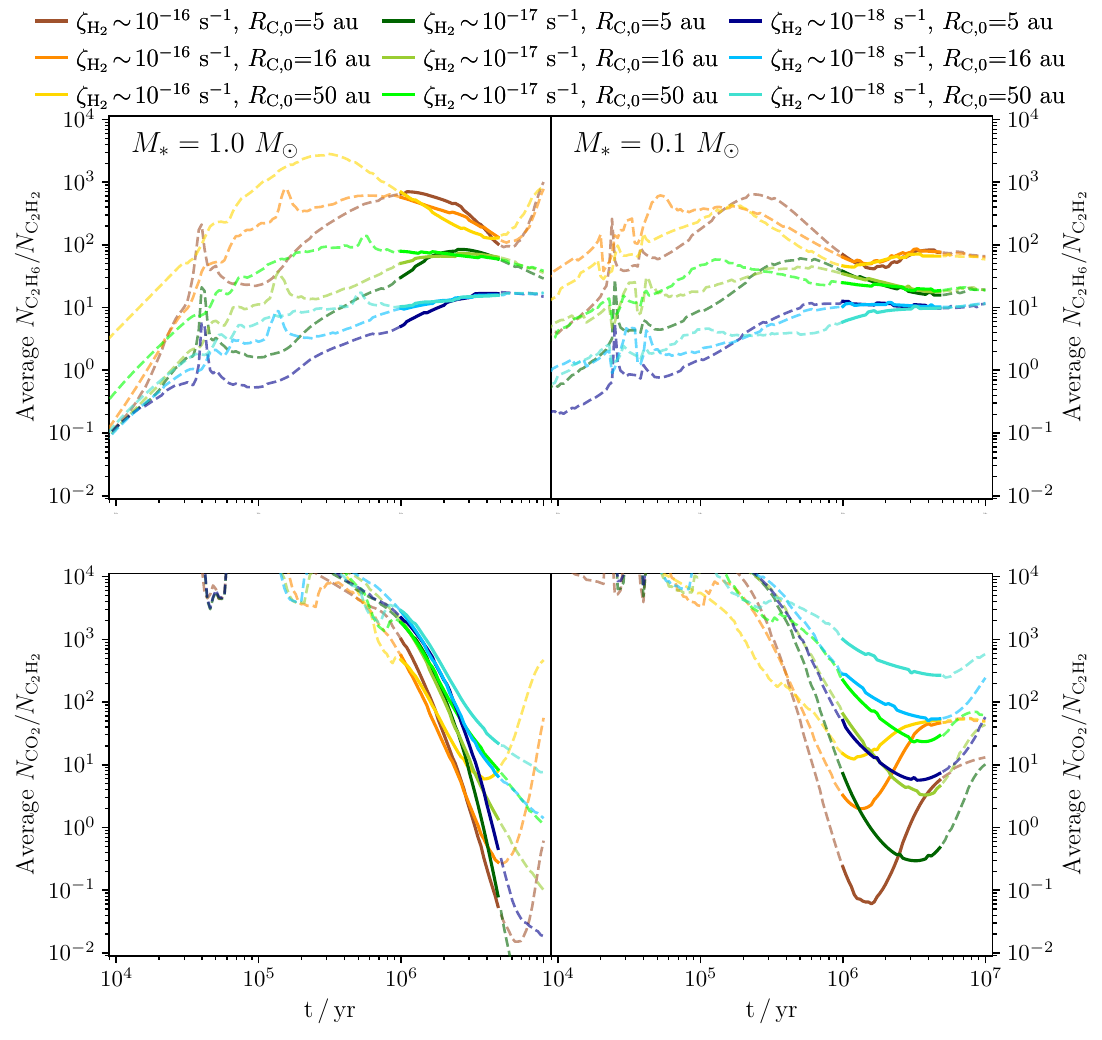}
    \vspace{-8pt}
    \caption{
    Evolution of molecular gas-phase column density ratios (top: \ce{C2H6} and \ce{C2H2}, bottom: \ce{CO2} and \ce{C2H2}), averaged across the disc inside their respective snowlines.
    }
    \label{fig:molecules}
\end{figure*}
Figure \ref{fig:hydrocarbons} focuses on the inner-disc gas-phase C/O ratio, rather than individual molecules, as significant chemical reprocessing may happen in the observable upper layers, with the complexity of the products depending on the relative roles of UV or X-ray radiation \citep{Walsh_2015,Raul_2025}.
For reference, however, \ref{fig:molecules} shows the evolution of averaged \ce{CO2}/\ce{C2H2} and \ce{C2H6}/\ce{C2H2} column density ratios for these models. 

The column density ratios were approximated as the ratio of area-averaged observable column densities 
\begin{equation}
    \frac{N_x}{N_y} = \frac{\langle N_{x,\rm obs}\rangle}{\langle N_{y,\rm obs}\rangle}
\end{equation}
where
\begin{equation}
    \langle N_{x,\rm obs}\rangle = \frac{\int_0^{R_{\rm snow}} N_x f_{\rm obs} 2\pi{\rm d}R}{\pi R_{\rm snow}^2} 
\end{equation}

Over time, as more \ce{C2H2} is created and as the less-volatile \ce{CO2} advects onto the star on a shorter timescale than \ce{C2H2}, the \ce{CO2}/\ce{C2H2} falls. This is in line with the tentative observational trends, where \citet{Arabhavi_2025a} find that VLMSs in the older Upper Sco star forming region have lower \ce{CO2}/\ce{C2H2} flux ratios. Moreover, \citet{Long_2025} note an average increase in the $N_{\ce{C2H2}}/N_{\ce{CO2}}$ column density ratio over time - with the extremes being the young \ce{H2O}-rich VLMS Sz\,114 and the 34 Myr old J0446B - which they attributed to an increase in C/O. The increase we see is not driven by the fact that equilibrium chemistry forms more \ce{C2H2} and less \ce{CO2} at higher C/O as in thermochemical models, but simply as a result of the secular evolution of the disc as the molecules are formed and transported. At late times, the hydrocarbon-rich gas starts to accrete onto the star to be replaced with C-depleted gas, and the \ce{CO2}/\ce{C2H2} ratio rises again.

The \ce{C2H6}/\ce{C2H2} ratio is strongly dependent on the ionization rate, and rather weakly on the disc size. The values are similar for both stellar masses at a given ionization rate and in all cases are very high, as hydrogenation is very efficient and as in our simplified approach, \ce{C2H6} acts as a sink. Photodissociation reactions of \ce{C2H6} (and \ce{C2H4}) could help reverse this balance as could the formation of more complex hydrocarbons or organic molecules via other reactions that were not considered.

\subsection{Impact of carbonaceous grain erosion}
Carbonaceous grains can be eroded by various processes, as outlined in the introduction. 
While most of these need high temperatures to become efficient and are thus restricted to the inner disc, previous modelling suggests that the destruction of carbonaceous grains by UV at large distances from the star could be responsible for the high C/O observed there in some discs \citep{Bergin_2016,Anderson_2017,Bosman_2021b}. As the focus of this work is the connection between the outer disc chemical processing and the inner disc, we focus on whether this `UV photolysis' would affect our results, or if the inner-disc C/O can only be raised by grain erosion if it happens in situ.

We follow the approach used by \citet{Anderson_2017,Bosman_2021b} who based their rates of destruction by UV on measurements from photolysis experiments of hydrogenated amorphous carbon grains, in which \ce{CH4} was the primary product, expressed in terms of the UV photon flux $F_{\rm UV} = 10^8 G_0$ and the yield of C atoms per photon $Y_{\rm UV}=8\times10^{-4}$ \citep{Alata_2014,Alata_2015}:
\begin{equation}
    k_{\rm phot}=\frac{\sigma Y_{\rm UV} F_{\rm UV}}{N_{\rm C}}
    ,
\end{equation}
where $\sigma=\pi a^2$ is the geometric cross-section of the grains, and $N_{\rm C} = (4\pi\rho_s a^3)/(3\mu_{\rm C} m_H)$ is the number of carbon atoms per grain where $\mu_{\rm C}=12$ is the relative atomic mass of C and $\rho_s=2.24\,\mathrm{g\,cm^{-3}}$ is the bulk density of carbonaceous material.
These factors combine to give $(3\mu_{\rm C} m_H)/(4a\rho_s)$ showing that even if the large grains are exposed to the UV (e.g. in the outer disc), their low surface area-to-volume ratio means they are inefficiently eroded. We therefore consider only small grains (assumed to be $0.1\,\mu\mathrm{m}$) for our calculation.

\label{appendix:photolysis}
\begin{figure*}[t]
    \centering
    \includegraphics[width=\linewidth]{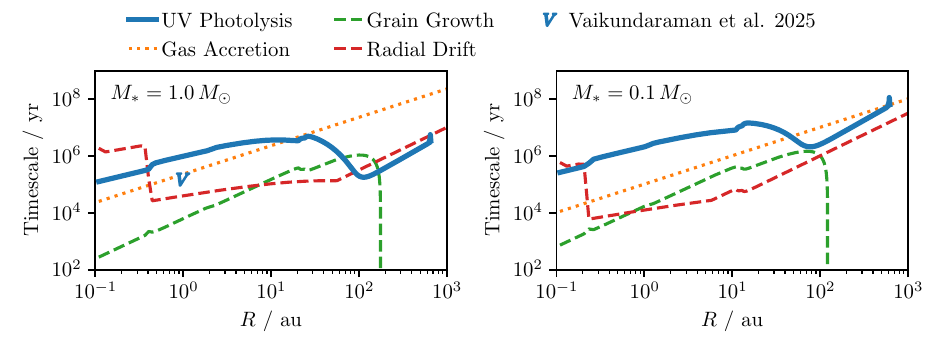}
    \vspace{-24pt}
    \caption{Comparison of the timescales at $t=0\,\mathrm{yr}$ around a $1.0\,M_\odot$ star (left) and $0.1\,M_\odot$ star (right) for the erosion of carbonaceous grains by UV photolysis (blue, solid) with the timescales of gas accretion (orange, dotted; see Equation \ref{eq:taccmol}), dust grain growth (green, dashed) and radial drift (red, dashed). The carbon depletion timescale at 1\,au in the `fiducialfc025' model from \citet{Vaikundaraman_2025} is shown as the blue V.}
    \label{fig:UVtimescales}
\end{figure*}

The disc evolution code is a 1D, vertically-integrated, code, which tracks the mass fractions $\epsilon_i = \Sigma_i/\Sigma$. We express the rate of change of the carbonaceous grain mass fraction as
\begin{align}
    \frac{{\rm d}\epsilon_i}{{\rm d}t} &= \frac{1}{\Sigma} \frac{{\rm d} \Sigma_i}{{\rm d}t} \\
    &=\frac{2}{\Sigma} \int_0^\infty \frac{{\rm d} \rho_i(z)}{{\rm d}t} {\rm d}z \\
    &=\frac{2}{\Sigma} \frac{\sigma Y_{\rm UV}}{N_{\rm C}} \int_0^\infty \rho_i(z) F_{\rm UV, z} {\rm d}z
    ,
\end{align}
We assume the UV is attenuated by the vertical column of gas as $F_{\rm UV, z}=F_{\rm UV,0}e^{-\tau_{\rm UV, z}}$ where the optical depth to UV $\tau_{\rm UV,z} = \sigma_H (\epsilon_d/0.01) \int_z^\infty{n_H(z') {\rm d}z'}$ for UV cross section per hydrogen atom $\sigma_{\rm H}=2.6\times10^{-21}\,\mathrm{cm^{-2}}$ \citep{Facchini_2016}.
By writing $\mu_i m_H n_i = \rho_i = \epsilon_i \rho  = \epsilon_i \mu m_H n$ assuming the composition is vertically uniform, and taking $n_{\rm H} = 2 n_{\rm H_2}$, we have 
$\tau_{\rm UV,z} = \mu \sigma_H (\epsilon_d/0.01) \epsilon_{\rm H_2} \int_z^\infty{n(z') {\rm d}z'} = \sigma_{\rm eff} N_z$ where $\sigma_{\rm eff} = \mu \sigma_H (\epsilon_d/0.01) \epsilon_{\rm H_2}$ and $N_z = \int_z^\infty{n(z') {\rm d}z'}$. Noting that ${n(z) {\rm d}z} = - {\rm d}N_{z}$, we can then write 
\begin{align}
    \frac{{\rm d}\epsilon_i}{{\rm d}t} &= \frac{2\epsilon_i \mu m_H}{\Sigma} \frac{\sigma Y_{\rm UV}}{N_{\rm C}} \int_{z=\infty}^{z=0} F_{\rm UV, 0} e^{-\sigma_{\rm eff} N_z} {\rm d}N_z \\
    &=\epsilon_i \frac{\sigma Y_{\rm UV}}{N_{\rm C}} \frac{1-e^{-\tau_{\rm UV, 0}}}{\tau_{\rm UV, 0}}
    ,
\end{align}
i.e. the vertically-averaged rate is modulated by a factor $(1-e^{-\tau_{\rm UV, 0}})/\tau_{\rm UV, 0}$.
By using the vertically averaged abundance and rate, we assume that the material is well-mixed vertically, which is reasonable for our purposes as the timescale for vertical mixing is shorter than the timescale for accretion from the outer disc to the inner disc by a factor $h^2$.
This approach is in contrast to the works above which are static thermochemical models with a vertical C/O gradient that only rises above 1 in the surface layers that are optically thin to the UV.
As any grain effectively spends only $(1-e^{-\tau_{\rm UV, 0}})/\tau_{\rm UV, 0}$ of its time in the optically thin layers due to vertical mixing, the impact on raising C/O is essentially diluted by this factor under our assumptions.
We note that this `unrestricted photolysis regime' \citep{Klarmann_2018,Vaikundaraman_2025} is a best case scenario: the average erosion may be less efficient if the residence time of individual grains in the upper layers is longer than their destruction timescale $1/k_{\rm phot}$, such that erosion becomes limited by the rate of vertical mixing of fresh grains to the surface \citep{Anderson_2017,Klarmann_2018}. 
In order to fully deplete the carbon from the grains, efficient collisions are also required to recycle material from larger grains into smaller grains that are able to reach the exposed surface \citep{Vaikundaraman_2025}.

As we argue that photolysis is only relevant for the small dust grains, and as these are the only grains which can reside in the upper layers where the UV can penetrate, we scale the rate with the fraction of dust in small grains which, in the two population model, is $f_{\rm small}=0.03-0.25$ of the total grain mass \citep[in the drift limit and fragmentation limit respectively][]{Birnstiel_2012}. This factor is assumed to modify both the optical depth to UV $\tau_{\rm UV, 0}$ and the final rate ${\rm d}\epsilon_i/{\rm d}t$, though in the limit of large $\tau_{\rm UV, 0}$, the two contributions will cancel, as, essentially, the UV penetrates (and can thus erode) a fixed number of grains

The incident UV field $F_{\rm UV, 0}$ is given by a $12000,\mathrm{K}$ blackbody with luminosity $L_{\rm FUV, 0}$ equal to the accretion luminosity:
\begin{equation}
    L_{\rm FUV, 0} = f_{\rm UV}(12000\,\mathrm{K}) \frac{0.8 G M_* \dot{M}}{R_*}
    ,
\end{equation}
where $f_{\rm FUV}(12000\,\mathrm{K})=0.245$ is the fraction of a $12000\,\mathrm{K}$ blackbody's luminosity radiated at FUV energies.
The UV radiation (in Habing units) intercepted per unit area  of the disc is then
\begin{equation}
    G_0 = \frac{L_{\rm acc}}{1.63\times10^{-3}\,\mathrm{erg\,s^{-1}\,cm^{-2}}} \frac{\phi}{4\pi r^2} 
    ,
\end{equation}
where $\phi \sim\!h$ is the flaring angle of the disc surface.

The model is also included in Fig. \ref{fig:onebyone} as the model involving C-grains (purple) and in the final row of Table \ref{tab:onebyone}. A very small (sub-percentage) increase is seen for the VLMS model.
This can be understood by comparing the grain destruction timescale to the timescales of the dust and gas evolution (Fig. \ref{fig:UVtimescales}).
Although the timescales for grain destruction are shorter than the typical disc lifetime, in these models, over much of the disc, the dust can grow to large grains, drift, and accrete onto the star faster than this timescale. Thus a very small fraction of the dust is eroded before it is accreted onto the star, and so although there was initially a large reservoir of refractory grains, it is not very effectively liberated for either $M_*$. This is in agreement with the results of \citet{Klarmann_2018}.
We note that our UV photolysis timescale estimate is somewhat longer, but of the same order of magnitude as, that of \citet{Vaikundaraman_2025}: the differences could be due to differences in UV luminosity and scattering or grain size distributions, compositions, and structure, but nevertheless our conclusions would still hold.
Models with less efficient growth and drift might allow a larger fraction of the carbonaceous grains to be destroyed and to contribute to enriching the gas, for example if they are retained at a dust trap (similar to the scenario for \ce{CH4} production from \ce{CH3OH} in Section \ref{sec:results_traps}).

\FloatBarrier

\section{Observational outliers}
So far, two discs observed by MIRI-MRS stand out as outliers to the trend of C-rich VLMS discs and O-rich TTS discs.

DoAr\,33 has a C-rich inner disc with C/O$\,=\!2-4$ \citep{Colmenares_2024} around 1 $1.1\,M_\odot$ star in the $0.3-6\,\mathrm{Myr}$ $\rho$\,Ophiuchi region. Explaining this source, on these timescales, via transport of outer disc material would require an ionization rate elevated by a factor $\sim\!10$ above the galactic rate. This depletes O sufficiently, while C remains somewhat enhanced inside the CO snowline following sublimation of drifting ices, to reproduce the inferred abundance for this disc (without additional mechanisms e.g. erosion of carbonaceous grains).

Conversely, Sz\,114, a \ce{H2O}-rich disc around a VLMS \cite{Xie_2023}, has an estimated $R_{\rm C}=18.2\,\mathrm{au}$ \citep{Trapman_2023}, making it larger than average, consistent with its status as one of the brightest discs at mm wavelengths around a mid-to-late M-type star in the Lupus star-forming region \citep[with mm flux being correlated with mm disc size,][]{Tripathi_2017,Andrews_2018b,Kurtovic_2021}. The large size of Sz\,114 could play a role in its relatively C-poor chemistry, as explored in Section \ref{sec:results_size}. Alternatively, Sz\,114 may simply me too young to have yet crossed to C/O$>1$ \citep{Xie_2023,Long_2025}. In our models, this could happen if it is $\lesssim2\,\mathrm{Myr}$, consistent with Lupus' age range. Finally, the \ce{H2O}-rich phase could be prolonged past that time if the disc hosts leakier traps \citep[e.g.][]{Mah_2024} than those we explored in Section \ref{sec:results_traps} \& Appendix \ref{appendix:traps} and so long as gas-phase \ce{CH4} is not produced there (e.g. if the traps are cold enough for it to remain frozen out, or the photolysis of \ce{CH3OH} is less effective than assumed).

\FloatBarrier

\section{Further examples of models with traps}
\label{appendix:traps}
\begin{figure}[h]
    \centering
    \includegraphics[width=\linewidth]{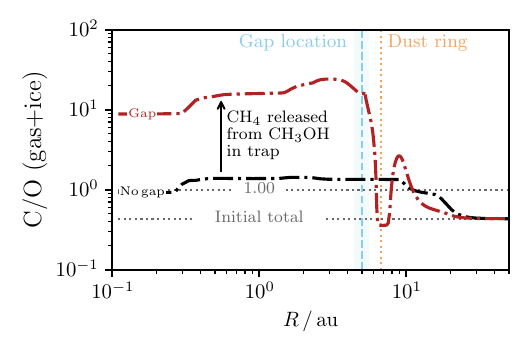}
    \vspace{-24pt}
    \caption{C/O as a function of $R$ at 2.5\,Myr for models with no trap (black) or a warm trap (red) in a 50 au disc around a $0.1\,M_\odot$ star with an ionization rate of $1.3\times10^{-17}\,\mathrm{s^{-1}}$. Vertical lines mark the location of the minimum gas density ('Gap location') and the pressure maximum where dust accumulates ('Dust ring').
    }
    \label{fig:COvsR}
\end{figure}
Figure \ref{fig:COvsR} shows the C/O ratio as a function of radius for the warm trap model in a 50 au disc around a $0.1\,M_\odot$ star as explored in Section \ref{sec:results_traps}. Since the C/O gradient is similar to the case inferred for J1605, we plot this model at 2.5\,Myr, the dynamical traceback age of the $\beta$\,Sco subregion of Upper Sco where J1605 is located \citep{MiretRoig_2022}.

\begin{figure*}
    \centering
    \includegraphics[width=0.75\linewidth]{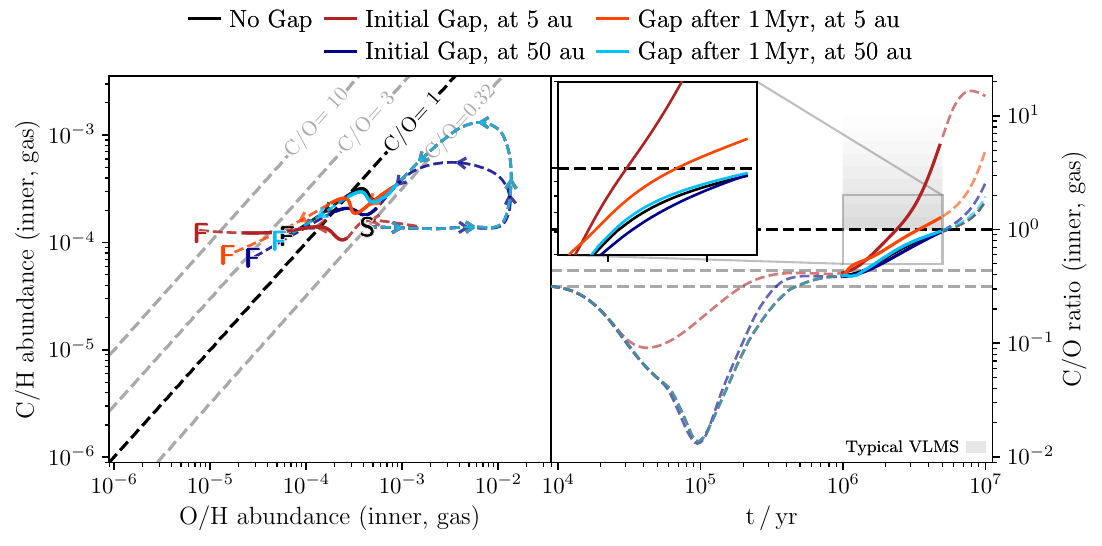}
    \vspace{-8pt}
    \caption{As Fig. \ref{fig:trapParams} but for $1\,M_\odot$.}
    \label{fig:trapParams_TT}
\end{figure*}

\begin{figure*}
    \centering
    \includegraphics[width=0.75\linewidth]{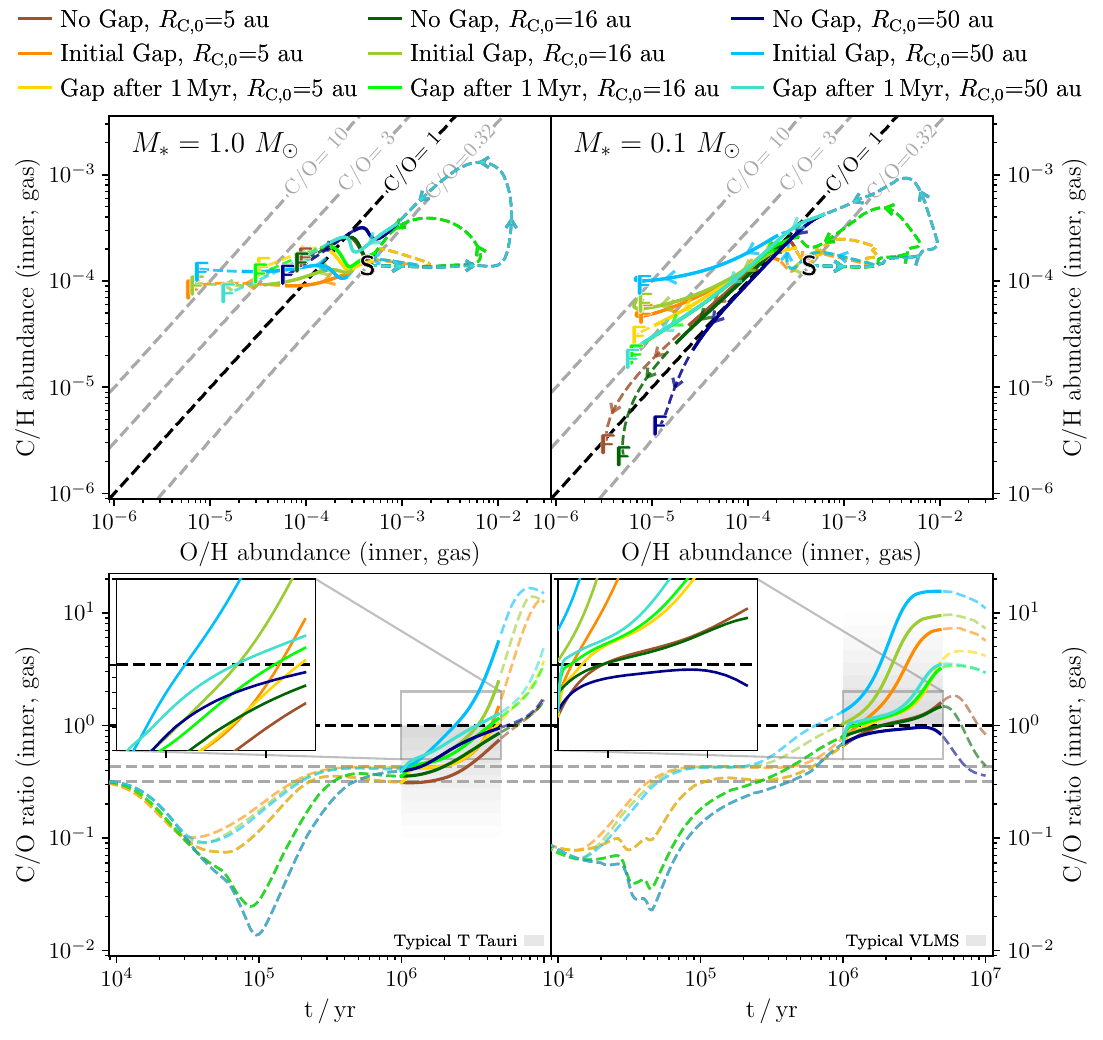}
    \vspace{-8pt}
    \caption{As Fig. \ref{fig:trapParams} and \ref{fig:trapParams_TT} but comparing models with and without warm dust traps for a range of initial disc radii.}
    \label{fig:traps}
\end{figure*}

Figure \ref{fig:trapParams_TT} shows equivalent warm and cold-trap models to Fig. \ref{fig:trapParams} for $1.0\,M_{\odot}$ stars and Fig. \ref{fig:traps} shows warm-trap models for all initial radii.
A similar effect is seen for all $R_{\rm C,0}$ investigated, for example with warm traps photodissociating \ce{CH3OH} and leaking \ce{CH4} into the inner disc. Most notably, more material starts outside the trap for larger discs and therefore in these cases, there is a larger trapped \ce{CH3OH} reservoir which can sustain a higher C/H and C/O. 

\FloatBarrier

\section{Comparing the inner and outer disc}

\label{appendix:outer}
\begin{figure*}
    \centering
    \includegraphics[width=\linewidth]{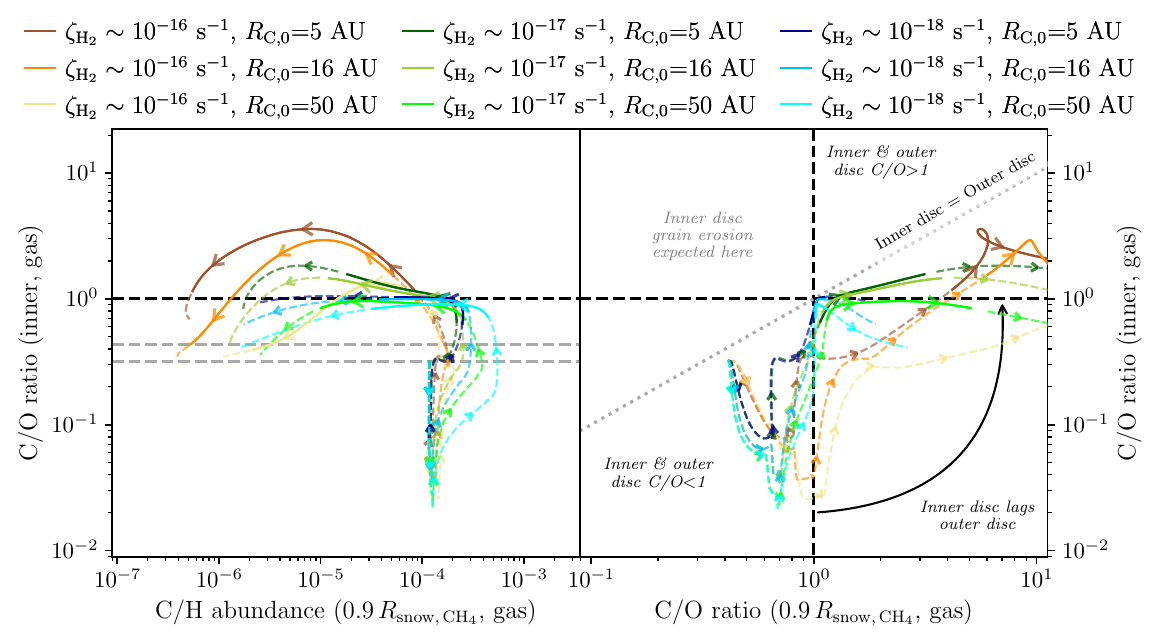}    
    \vspace{-8pt}
    \caption{Relationship between the inner disc C/O (measured at the 0.1 au inner boundary of the models) and the outer disc (measured at 90\% of the \ce{CH4} snowline) C abundance (left) and C/O (right) for discs around VLMSs.}
    \label{fig:innerouter}
\end{figure*}

Figure \ref{fig:innerouter} compares the inner and outer disc chemistry for the model grid with hydrocarbon formation. As before, the inner disc is measured at the inner boundary of the model in order to represent the hottest gas in the disc, since most lines are observed with excitation conditions of several 100s K. For the outer disc, the choice of is more arbitrary (as we are mostly looking to demonstrate the lag between the two), but since we are interested in the volatile C and O abundances, we need to measure at a location where both of the main volatile C-carriers - CO and \ce{CH4} - are in the gas phase and thus choose just inside the \ce{CH4} snowline. At this location, the C/H initially rises a little as \ce{CO} and \ce{CH4} ices desorb at their snowlines, but eventually starts to deplete by up to 2 orders of magnitude \citep[consistent with typical TTS disc measurements][]{Favre_2013,Kama_2016b,McClure_2016,Miotello_2017,Anderson_2022,Trapman_2022,PanequeCarreno_2025} in line with the gas outside the snowlines.

More interesting is the relationship between inner and outer disc C/O. Initially, we measure C/O$<1$ in both the inner and outer disc, due to the presence of \ce{O2} gas in the outer disc. The inner disc C/O initially drops quite rapidly as O-rich \ce{H2O} ice is deposited in the inner disc by drift. Conversely, the enhancement at the \ce{CH4} and \ce{CO} snowlines takes slightly longer to occur, as the dust grains take longer to grow and start to drift further out in the disc, and as a larger fraction of the molecules are already in the gas-phase any potential enhancement is weaker. Thus, the model tracks initially drop almost vertically on the plot of inner disc vs outer disc C/O (Fig. \ref{fig:innerouter}). As the O-rich gas starts to accrete, they reverse this course, the timescale for this accretion is very short due to the proximity of the \ce{H2O} snowline to the star, so the rise in inner disc C/O happens faster than that in the outer disc. Nevertheless, the discs do cross to C/O in the outer disc before they can do so in the inner disc, as once the remnants of the initial drift phase are accreted, the chemistry is driven by the inward spreading of the outer-disc gas. Moreover, the inner disc C/O is kept somewhat lower than around the \ce{CH4} snowline due to the sublimation of O-rich ices between those locations, especially \ce{H2O} (Fig. \ref{fig:water}). Thus, we see all the model tracks pass through the lower right quadrant of Fig. \ref{fig:innerouter}. Eventually, the inner disc may also reach C/O$>1$ and then the discs move into the upper right of the plot, but remain below the line where the inner and outer disc C/O are equal, though discs with sufficient resupply of \ce{H2O} may just remain in the bottom right (e.g. the lightest green track).

We note that if C/O is instead reached by the release of additional C in the inner disc, without O-depletion in the outer disc, we might expect discs to populate the upper left hand quadrant, where our models are never seen.
This may be a way to distinguish between the two proposed scenarios of high C/O due to inner disc grain erosion vs O-depleted gas.
For example, in the low-turbulence ($\alpha=10^{-4}$) models of \citep{Houge_2025b}, a strong, early pebble flux causes the irreversible thermal decomposition of refractory organics into \ce{C2H2} on drift timescales ($\lesssim0.1\,\mathrm{Myr}$), producing an early high C/O in the inner disc, while the composition of the outer disc has barely evolved (chemical reactions of CO are neglected in their work). On longer timescales however, the diffusion of this C-rich gas as far as the \ce{C2H2} snowline raises C/O above 1 outside the \ce{H2O} snowline and creates a longer-lived volatile C reservoir that can resupply the inner disc on longer timescales once the initial C-rich reservoir accretes onto the star. Thus, we might expect these models to approach similar conditions to ours at late times, but future work should compare the two directly.
Finally, we also caution that Fig. \ref{fig:COvsR} shows an alternative way to achieve a higher C/O in the inner disc than the outer disc, but we do find that in the plane shown in Fig. \ref{fig:innerouter}, our models stay in the upper right as the \ce{CH4} gas produced can also diffuse outwards towards its snowline.

\end{appendix}

\end{document}